\documentclass[twocolumn]{aastex62}

\newcommand{\Ni}{\ensuremath{^{56}\mathrm{Ni}}}

\newcommand{\Mej}{\ensuremath{M_\mathrm{ej}}}

\newcommand{\Msun}{\ensuremath{\mathrm{M}_\odot}}

\newcommand{\kmps}{\ensuremath{\mathrm{km~s^{-1}}}}
\newcommand{\ej}{\ensuremath{\mathrm{ej}}}

\graphicspath{{./}{figures/}}

\shorttitle{Fallback and SLSNe}
\shortauthors{Moriya, Nicholl, \& Guillochon}

\begin{document}

\title{
Systematic investigation of the fallback accretion powered model for hydrogen-poor superluminous supernovae
}

\correspondingauthor{Takashi J. Moriya}
\email{takashi.moriya@nao.ac.jp}

\author[0000-0003-1169-1954]{Takashi J. Moriya}\thanks{NAOJ Fellow}
\affil{Division of Theoretical Astronomy, National Astronomical Observatory of Japan, National Institutes of Natural Sciences, 2-21-1 Osawa, Mitaka, Tokyo 181-8588, Japan}

\author[0000-0002-2555-3192]{Matt Nicholl}
\affil{Harvard-Smithsonian Center for Astrophysics, 60 Garden Street, Cambridge, Massachusetts 02138, USA}

\author[0000-0002-9809-8215]{James Guillochon}
\affil{Harvard-Smithsonian Center for Astrophysics, 60 Garden Street, Cambridge, Massachusetts 02138, USA}



\begin{abstract}
The energy liberated by fallback accretion has been suggested as a possible engine to power hydrogen-poor superluminous supernovae. We systematically investigate this model using the Bayesian light-curve fitting code \texttt{MOSFiT} (Modular Open Source Fitter for Transients), fitting the light curves of 37 hydrogen-poor superluminous supernovae assuming a fallback accretion central engine. We find that this model can yield good fits to their light curves, with a fit quality that rivals the popular magnetar engine models. Examining our derived parameters for the fallback model, we find the total energy requirements from the accretion disk are estimated to be $0.002-0.7~\Msun c^2$. If we adopt a typical conversion efficiency $\sim 10^{-3}$, the required mass to accrete is thus $2 - 700~\Msun$. Many superluminous supernovae, therefore, require an unrealistic accretion mass, and so only a fraction of these events could be powered by fallback accretion unless the true efficiency is much greater than our fiducial value.
The superluminous supernovae that require the smallest amounts of fallback mass still remain to be the fallback accretion powered supernova candidates, but they are difficult to be distinguished solely by their light curve properties.
\end{abstract}



\section{Introduction} \label{sec:introduction}
Superluminous supernovae (SLSNe) are the most intrinsically luminous supernovae (SNe) currently known (see \citealt{moriya2018slsnrev} for a review). Despite significant interest in studying these events, the power source of hydrogen-poor (Type~I) SLSNe\footnote{We call Type~I SLSNe simply as SLSNe in this paper. We do not discuss Type~II SLSNe that are mainly powered by the interaction \citep[e.g.,][]{moriya2013sn2006gy}.} \citep[e.g.,][]{quimby2011slsn} is still debated. Major suggested power sources are the nuclear decay of \Ni\ \citep[e.g.,][]{gal-yam2009sn2007bi,moriya2010sn2007bi,kozyreva2017rapidpisn}, the interaction between SN ejecta and dense circumstellar media (CSM) \citep[e.g.,][]{chevalier2011irwin,moriya2012dip,sorokina2016slsnicsm}, and prolonged heating by some sort of a central engine \citep[e.g.,][]{kasen2010mag,woosley2010mag,metzger2015magtran}. It is also possible that several energy sources are active at the same time in SLSNe \citep[e.g.,][]{chen2017lsq14mo,tolstov2017ppisn}.

Late-phase observations of SLSNe have revealed that their nebular phase spectra resemble those of broad-line Type~Ic SNe which are often associated with long gamma-ray bursts \citep[e.g.,][]{pastorello2010sn2010gx,nicholl2016sn2015bnneb,jerkstrand2017slsnneb}, and the host galaxies of these two classes are also similar \citep[e.g.,][]{lunnan2014slsnhost,perley2016slsnhost,chen2017toogoodrelation,schulze2016slsnhost,angus2016slsnhost}. The similarity of SLSNe to broad-line Type~Ic SNe implies the possible existence of a central engine in SLSNe. The most popular central engine proposed to account for the huge luminosity of SLSNe is a strongly magnetized, rapidly rotating neutron star (NS) called a ``magnetar''. Strong magnetic fields allow NSs to spin down quickly and convert their rotational energy to radiation \citep{shapiro1983}. Magnetars have long been suggested as a potential power source in SNe \citep[e.g.,][]{ostriker1971mag,maeda2007mag05bf}, and they are now intensively applied for SLSNe \citep[e.g.,][]{kasen2010mag,woosley2010mag,dessart2012magslsn,chatzopoulos2013anachi,inserra2013slsntail,nicholl2013ptf12damnature,metzger2015magtran,wang2015slsnniandmag,bersten2016asassn15lh,moriya2017mimic,liu2017manymag,yu2017manymag}. The recent statistical study by \citet{nicholl2017mosmag} that used the Bayesian light-curve fitting code \texttt{MOSFiT} \citep{guillochon2017mosfit} has found that magnetars with initial spin periods of $1.2-4$~ms and magnetic field strengths of $(0.2-1.8)\times 10^{14}~\mathrm{G}$ can explain the overall properties of SLSNe.

However, the central engines that can be activated in SNe are not limited to magnetars. One alternative is fallback accretion \citep[e.g.,][]{dexter2013kasen}. A part of the SN ejecta that does not acquire enough energy to escape eventually falls back \citep{michel1988acc,chevalier1989acc};
this ``fallback'' material would ultimately accrete onto the central compact remnant. Such an accretion can result in outflows that provide an additional energy to increase the energy and luminosity of the SN \citep{dexter2013kasen,moriya2018oglefall}. Short-term accretion onto a black hole caused by the direct collapse of a massive star may result in long gamma-ray bursts and their accompanying broad-line Type~Ic SNe (\citealt{woosley1993lgrb}, see \citealt{hayakawa2018collapser,barnes2017grbblic} for recent studies), while longer-term accretion may be able to power the excess luminosity seen in SLSNe \citep{dexter2013kasen}.
\citet{metzger2018magfall} recently point out that the fallback accretion could affect the energy input from magnetars as well.

In this paper, we systematically investigate the light curves (LCs) of SLSNe assuming fallback accretion as the central power source. By fitting SLSN LCs using \texttt{MOSFiT}, we study whether the fallback accretion powered model can satisfactorily reproduce the known SLSN LCs, and whether the required fallback accretion parameters are feasible. We first introduce our method in Section~\ref{sec:method}. The results of the fallback LC fitting is presented in Section~\ref{sec:results} and they are discussed in Section~\ref{sec:discussion}. We conclude this paper in Section~\ref{sec:conclusions}.

\section{Method}\label{sec:method}
\subsection{MOSFiT}
We use the {\tt Python}-based LC fitting code  to apply the fallback accretion model to SLSNe. We briefly summarise the procedure here, and defer to \citet{guillochon2017mosfit} and \citet{nicholl2017mosmag} for the details of the code. In short, \texttt{MOSFiT} adopts a Markov Chain Monte Carlo (MCMC) approach to fit multi-band LCs, and provides the posterior probability distributions for the free parameters in the model. We perform Maximum Likelihood analysis, since using a full Gaussian process regression was found to have negligible impact on the fit parameters in the case of a magnetar model applied to the same SLSN sample we use here \citep{nicholl2017mosmag}. As in \citet{nicholl2017mosmag}, we use the first 10,000 iterations in the MCMC algorithm to burn in the ensemble (see \citealt{guillochon2017mosfit} for the details of the burning process) and at least 25,000 total iterations are performed before judging whether the fitting is converged. The convergence is checked by evaluating the Potential Scale Reduction Factor (PSRF, \citealt{gelman1992psrf}). Reliable convergence is obtained when the PSRF is below 1.2 \citep{brooks1998psrf1p1} and we terminate our iterations when it is below 1.1.

The parameters and priors in the fitting procedure are essentially the same as in \citet{nicholl2017mosmag}, but there are some differences. The central engine is, of course, changed to the fallback accretion as described in the next section. We do not set the gamma-ray opacity as a free parameter. \citet{nicholl2017mosmag} consider that the magnetar spin-down energy is released in the form of gamma-rays, and take the gamma-ray opacity into account in heating the SN ejecta. In the fallback model, the source of energy is the kinetic energy of outflows, and dynamical interaction between the outflows and the SN ejecta provides the heat to power the LCs. During the fitting procedure, we assume the central energy input from fallback ($L_\mathrm{fallback}$) is 100\% thermalized, since the conversion efficiency is fully degenerate with $L_\mathrm{fallback}$.  We consider the importance of using a realistic efficiency in converting accretion to thermal energy in Section~\ref{sec:discussion}. All free parameters used in the fits are summarized in Table~\ref{tab:param}.

\begin{deluxetable}{lccccc}
\tablecaption{Free parameters and priors in our model. \label{tab:param}}
\tablehead{
\colhead{parameter} & \colhead{prior} & \colhead{min} & \colhead{max} & \colhead{mean}& \colhead{$\sigma$} 
}
\startdata
$L_1$ ($\mathrm{erg~s^{-1}}$) & log-flat & $10^{51}$ & $10^{57}$ & \nodata & \nodata \\
$t_\mathrm{tr}$ (day) & log-flat & $10^{-4}$ & 100 & \nodata & \nodata \\
\Mej\ (\Msun) & log-flat & 0.1 & 100 & \nodata & \nodata \\
$v_\mathrm{phot}$ ($10^3$~\kmps) & Gaussian & 1 & 30 & 1.47 & 4.3 \\
$\kappa_\ej$ ($\mathrm{cm^2~g^{-1}}$) & flat & 0.05 & 0.2 & \nodata & \nodata \\
$T_\mathrm{f}$ (1000~K) & Gaussian & 3 & 10 & 6 & 1 \\
$A_V$ (mag) & flat & 0 & 0.5 & \nodata & \nodata \\
$t_\mathrm{exp}$ (day) & flat & $-100$ & 0 & \nodata & \nodata \\
variance & log-flat & $10^{-3}$ & 100 & \nodata & \nodata
\enddata
\end{deluxetable}

\subsection{Central power input and constraints}
The fallback accretion rate eventually follows a power law $\propto t^{-\frac{5}{3}}$, where $t$ is time after explosion \citep{michel1988acc,chevalier1989acc}. Numerical fallback simulations show that the accretion rate is usually flat at earlier times \citep{zhang2008acc,dexter2013kasen}. The initial flat accretion rate could be related to the freefall accretion but the reverse shock would also alter the early accretion rate significantly \citep{zhang2008acc,dexter2013kasen}. The fallback accretion dominates at later times. We assume that the kinetic energy in the outflow from the accretion disk is proportional to the accretion rate and a fraction of the kinetic energy is thermalized by the inelastic collision between the disk outflow and SN ejecta. We therefore assume that the central energy input from the fallback accretion, proportional to the accretion rate, follows
\begin{equation}
L_\mathrm{fallback}(t) = 
\left\{ \begin{array}{ll}
       L_1 \left( \frac{t_\mathrm{tr}}{1~\mathrm{sec}}\right)^{-\frac{5}{3}} \equiv L_\mathrm{flat}  & (t<t_\mathrm{tr}) \\
       L_1 \left( \frac{t}{1~\mathrm{sec}}\right)^{-\frac{5}{3}}  & (t\ge t_\mathrm{tr})
  \end{array} \right. ,
\end{equation}
where $L_1$ is a constant and $t_\mathrm{tr}$ is a transition time from the initial flat accretion to the power-law accretion. Briefly, the energy input is assumed to be constant ($L_\mathrm{flat}$) until $t=t_\mathrm{tr}$ and then start to decline with $\propto t^{-\frac{5}{3}}$. In the fitting procedure, $L_1$ and $t_\mathrm{tr}$ are set as free parameters. We have also performed the fitting without $t_\mathrm{tr}$. In this case, the fallback accretion power is $L_1(t/1~\mathrm{sec})^{-5/3}$ from the beginning. We found the fitting results without $t_\mathrm{tr}$ are not much different from the results with $t_\mathrm{tr}$ and we show the results with $t_\mathrm{tr}$ in this paper.

The total input energy from the fallback accretion ($E_\mathrm{total}$) and the possible energy brought by neutrinos ($E_\nu\simeq 10^{51}~\mathrm{erg}$) are the only energy sources for the kinetic energy of SN ejecta in our model. Therefore, the total kinetic energy, roughly estimated as $E_\mathrm{K}\simeq M_\mathrm{ej}v_\mathrm{phot}^2/2$, should satisfy $E_\mathrm{K}<E_\mathrm{total}+E_\nu - E_\mathrm{rad}$, where $E_\mathrm{rad}$ is the total radiated energy. We constrain the parameters to vary within this condition. The constraint that the nebular phase should not be reached before 100~days as in \citet{nicholl2017mosmag} is also kept.

\subsection{SLSN sample}
Table~\ref{tab:slsnsample} shows the SLSN sample we use to fit the fallback accretion model. We have 37 SLSNe in our sample. This sample is taken from \citet{nicholl2017mosmag} and we refer to \citet{nicholl2017mosmag} for the selection criteria in our sample.

\begin{deluxetable}{lcl}
\tablecaption{SLSN sample\label{tab:slsnsample}}
\tablehead{
\colhead{name} & \colhead{redshift} & \colhead{reference} 
}
\startdata
SN 2005ap & 0.265  & \citet{quimby2007sn2005ap} \\
SN 2006oz & 0.376  & \citet{leloudas2012slsnbump} \\
SN 2007bi & 0.1279 & \citet{gal-yam2009sn2007bi} \\
               	& &	\citet{young2010sn2007bi} \\
SN 2010gx & 0.2297 & \citet{pastorello2010sn2010gx}, \\
                 & & \citet{quimby2011slsn} \\
SN 2011ke & 0.1428 & \citet{inserra2013slsntail} \\
SN 2011kf & 0.245  & \citet{inserra2013slsntail} \\
SN 2012il & 0.175  & \citet{inserra2013slsntail} \\
SN 2013dg & 0.265  & \citet{nicholl2014pesstoslsn} \\
SN 2013hy & 0.663  & \citet{papadopoulos2015decamslsn} \\
SN 2015bn & 0.1136 & \citet{nicholl2016sn2015bn,nicholl2016sn2015bnneb} \\
PTF09atu  & 0.5015 & \citet{quimby2011slsn} \\
PTF09cnd  & 0.2584 & \citet{quimby2011slsn} \\
PTF09cwl  & 0.3499 & \citet{quimby2011slsn} \\
PTF10hgi  & 0.0987 & \citet{inserra2013slsntail} \\
PTF11rks  & 0.1924 & \citet{inserra2013slsntail} \\
PTF12dam  & 0.1073 & \citet{nicholl2013ptf12damnature} \\
                 & & \citet{chen2015ptf12damhost} \\
                 & & \citet{vreeswijk2017earlybump} \\
iPTF13ajg & 0.740  & \citet{vreeswijk2014iptf13ajg} \\
iPTF13dcc & 0.5015 & \citet{vreeswijk2017earlybump} \\
iPTF13ehe & 0.3434 & \citet{yan2015iptf13ehe} \\
iPTF16bad & 0.2467 & \citet{yab2017slsnlatehalpha} \\
PS1-10ahf & 1.1    & \citet{mccrum2015slsnpanrate} \\
PS1-10awh & 0.908  & \citet{chomiuk2011panslsn} \\
PS1-10bzj & 0.650  & \citet{lunnan2013ps110bzjslsnandhost} \\
PS1-10ky  & 0.956  & \citet{chomiuk2011panslsn} \\
PS1-10pm  & 1.206  & \citet{mccrum2015slsnpanrate} \\
PS1-11ap  & 0.524  & \citet{mccrum2014slsnps111ap} \\
PS1-11bam & 1.565  & \citet{berger2012slsnism} \\
PS1-14bj  & 0.5215 & \citet{lunnan2016ps114bjlonghpoorslsn} \\
LSQ12dlf  & 0.255  & \citet{nicholl2014pesstoslsn} \\
LSQ14mo   & 0.253  & \citet{chen2017lsq14mo} \\
LSQ14bdq  & 0.345  & \citet{nicholl2015lsq14bdqclearbump} \\
Gaia16apd & 0.102  & \citet{nicholl2017gaia16apd} \\
                 & & \citet{yan2017gaia16apduv} \\
                 & & \citet{kangas2017gaia16apd} \\
DES14X3taz& 0.608  & \citet{smith2016des14x3taz} \\
SCP-06F6  & 1.189  & \citet{barbary2009scp06f6} \\
SNLS06D4eu& 1.588  & \citet{howell2013slsn} \\
SNLS07D2bv& 1.50   & \citet{howell2013slsn} \\
SSS120810 & 0.156  & \citet{nicholl2014pesstoslsn} \\
\enddata
\end{deluxetable}

\begin{figure*}
\gridline{\fig{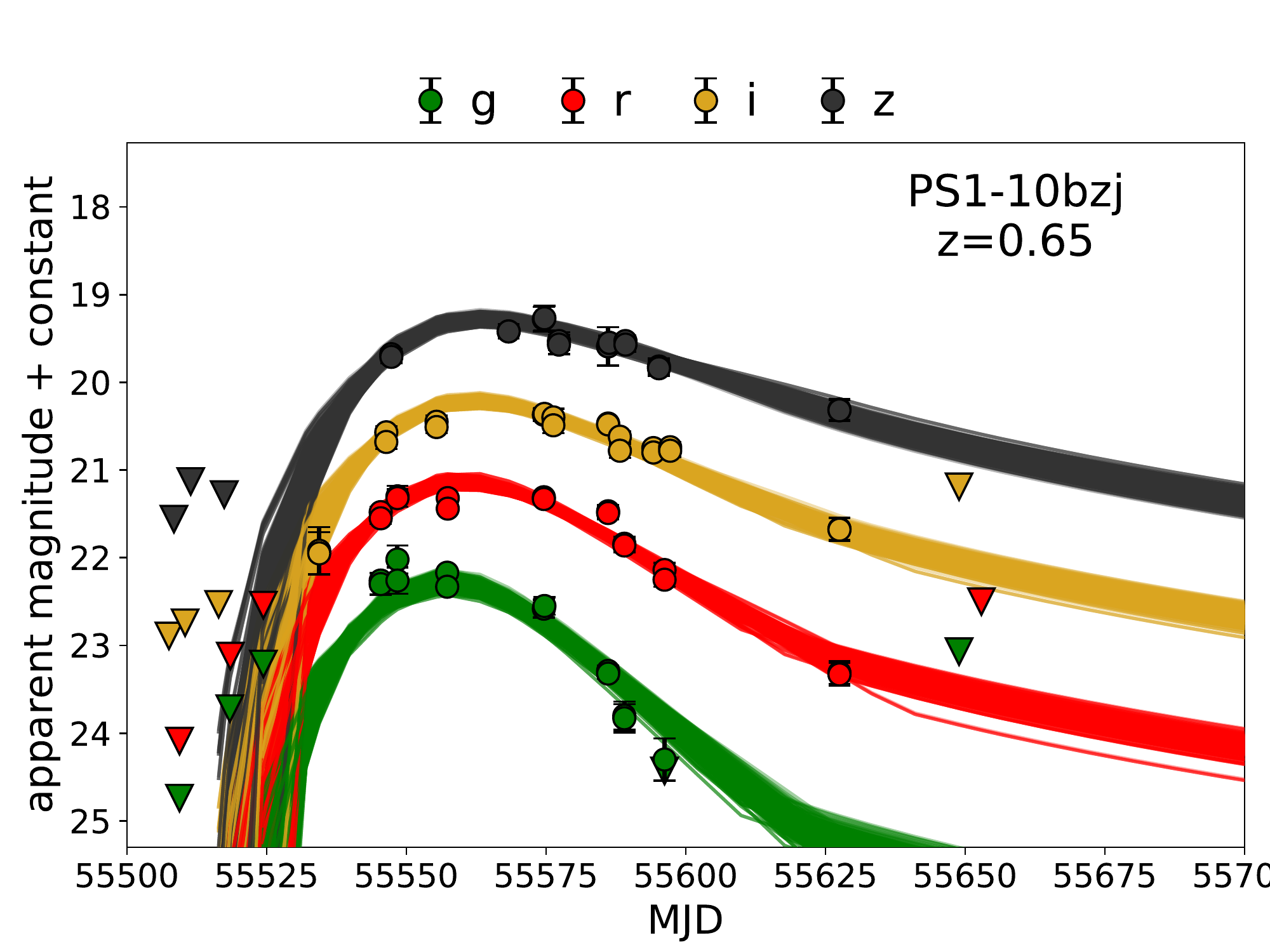}{0.33\textwidth}{}
          \fig{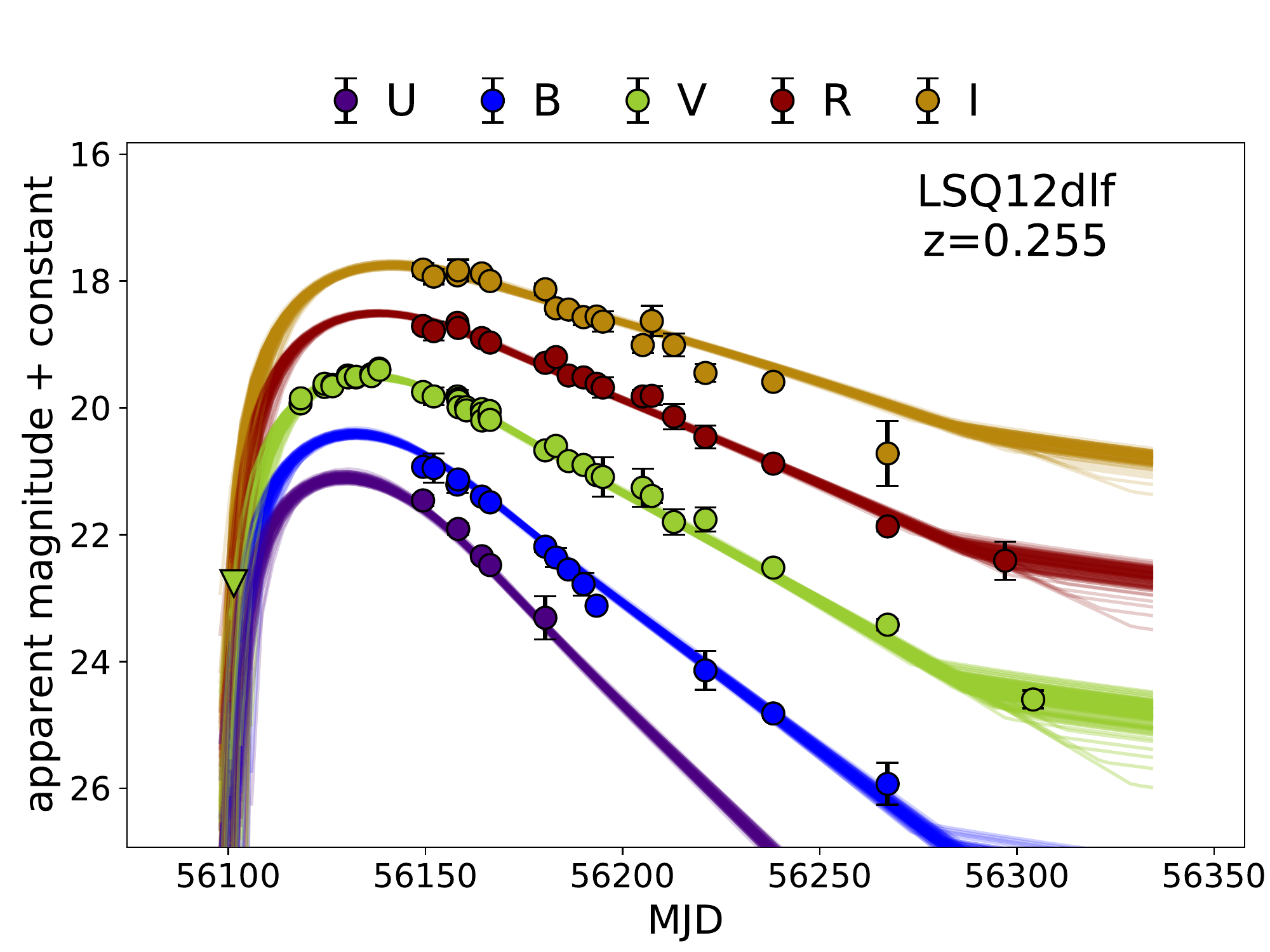}{0.33\textwidth}{}
          \fig{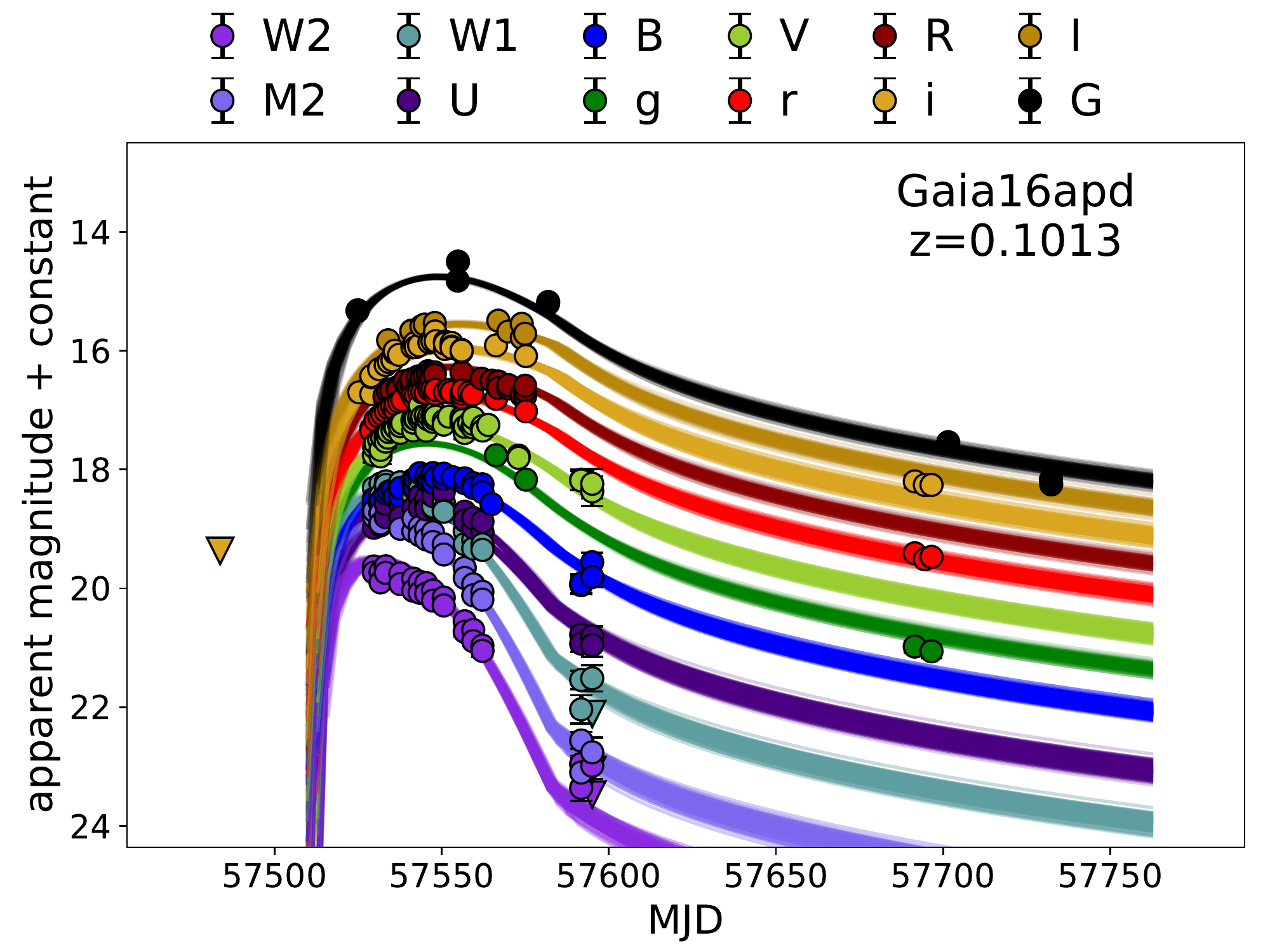}{0.33\textwidth}{}
          }
\gridline{\fig{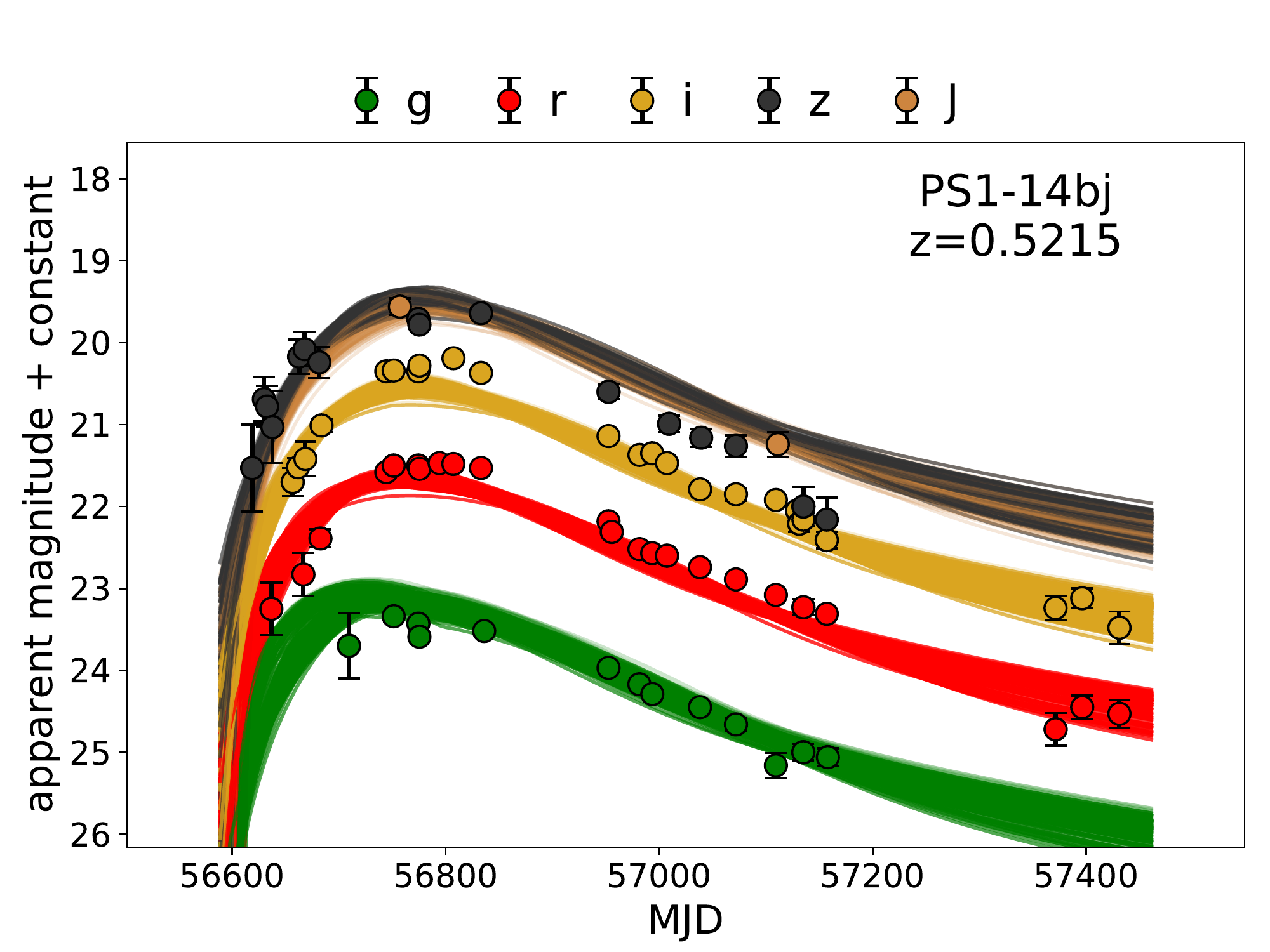}{0.33\textwidth}{}
          \fig{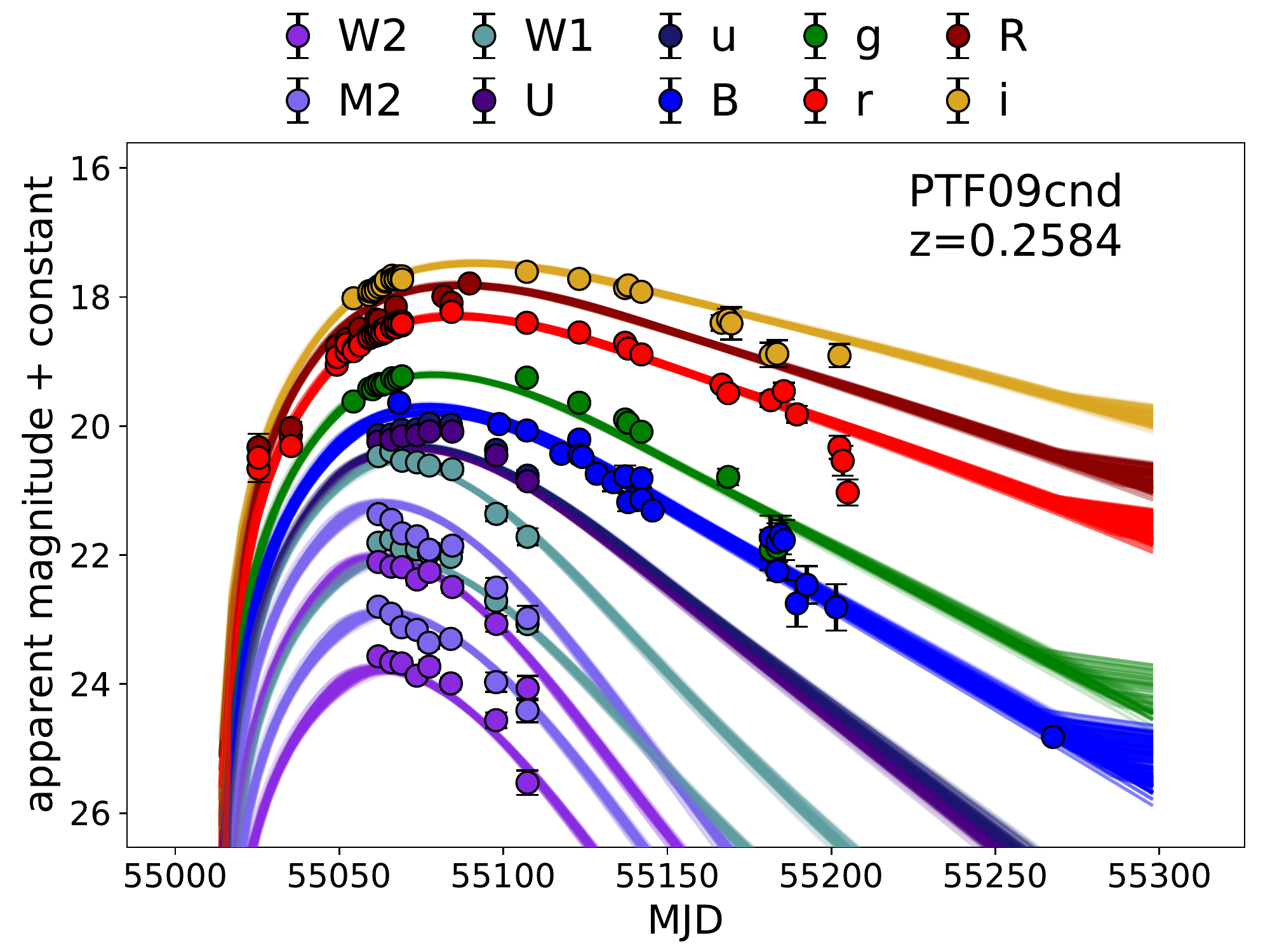}{0.33\textwidth}{}
          \fig{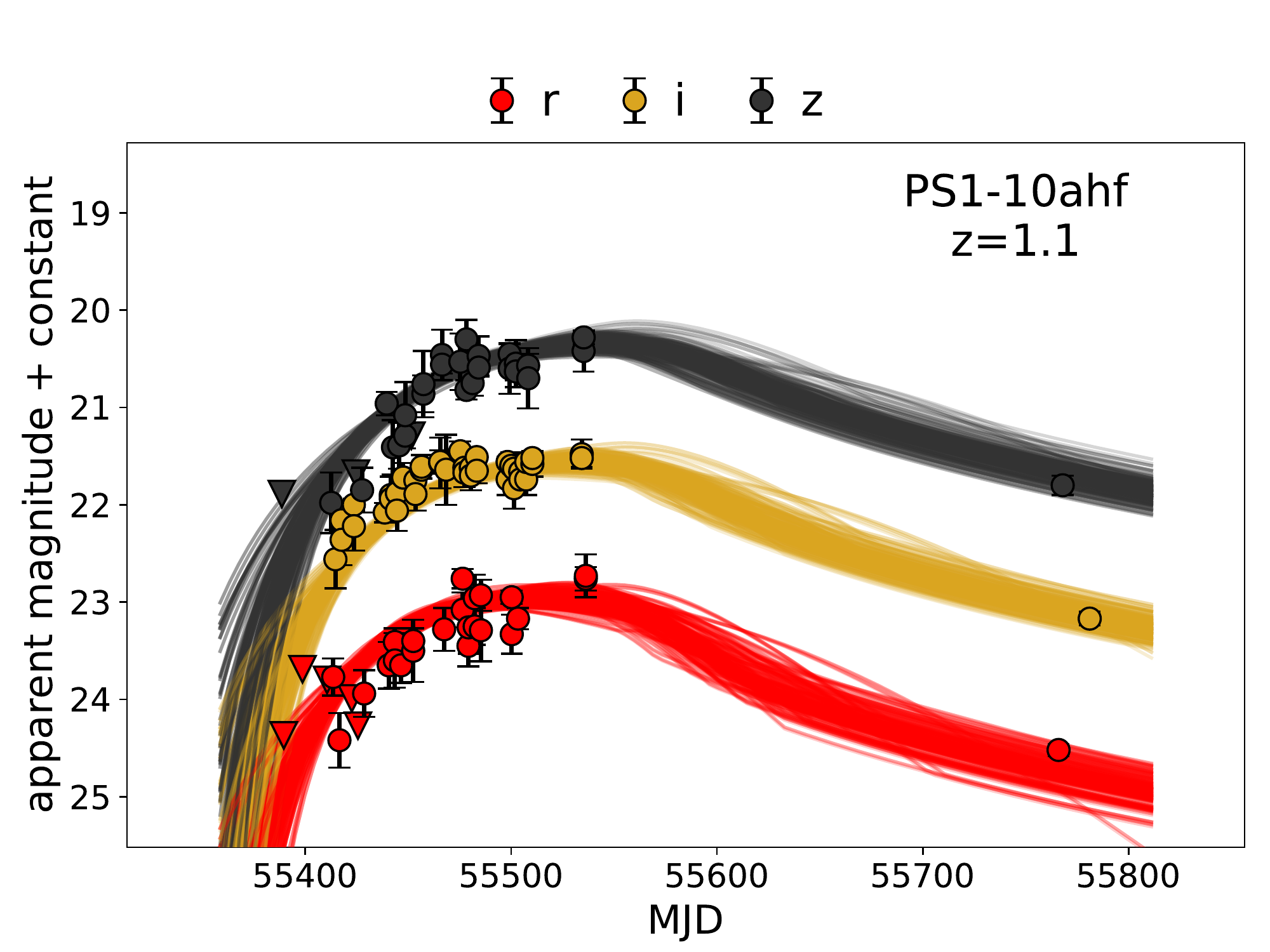}{0.33\textwidth}{}
          }
\gridline{\fig{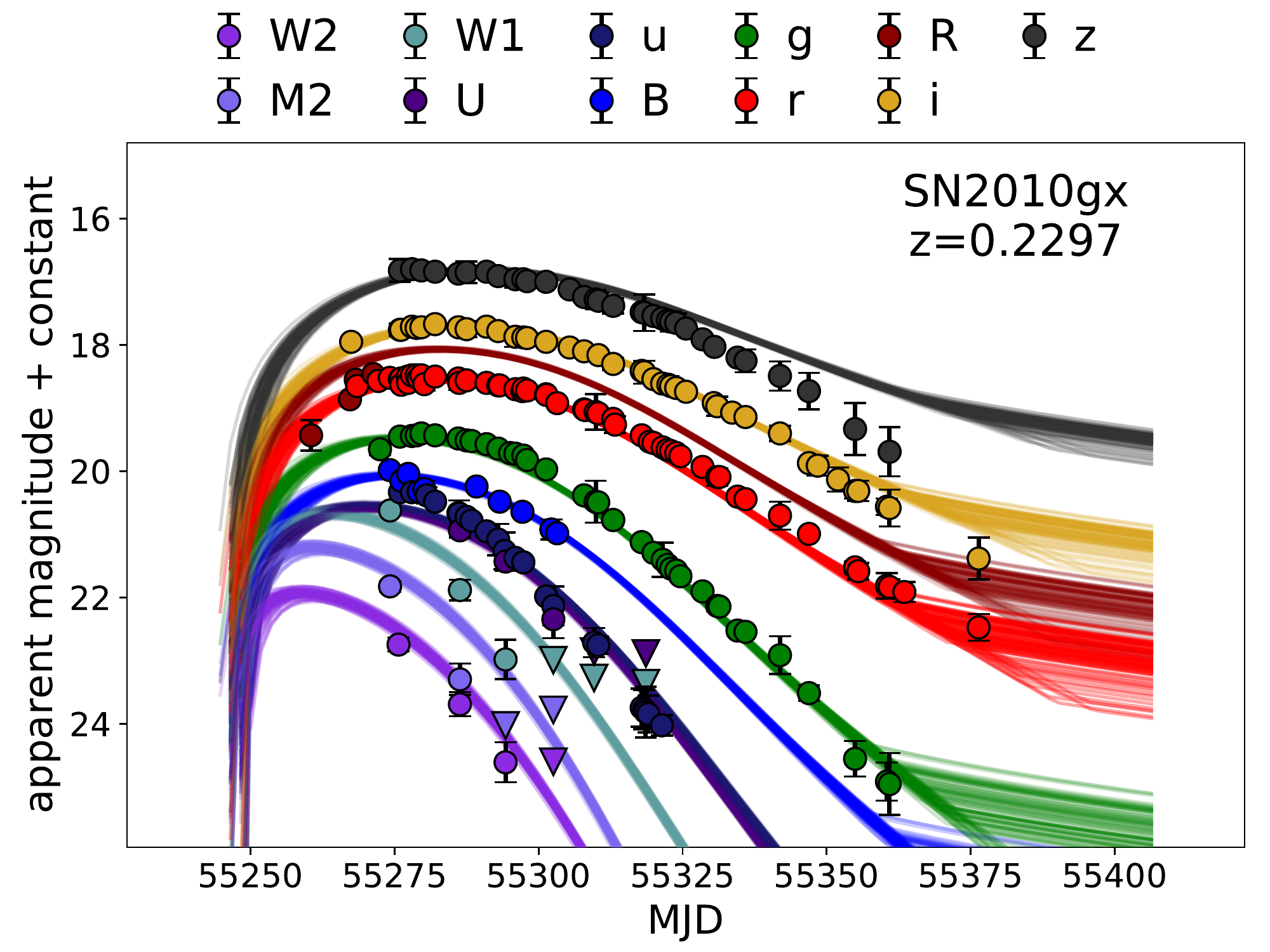}{0.33\textwidth}{}
          \fig{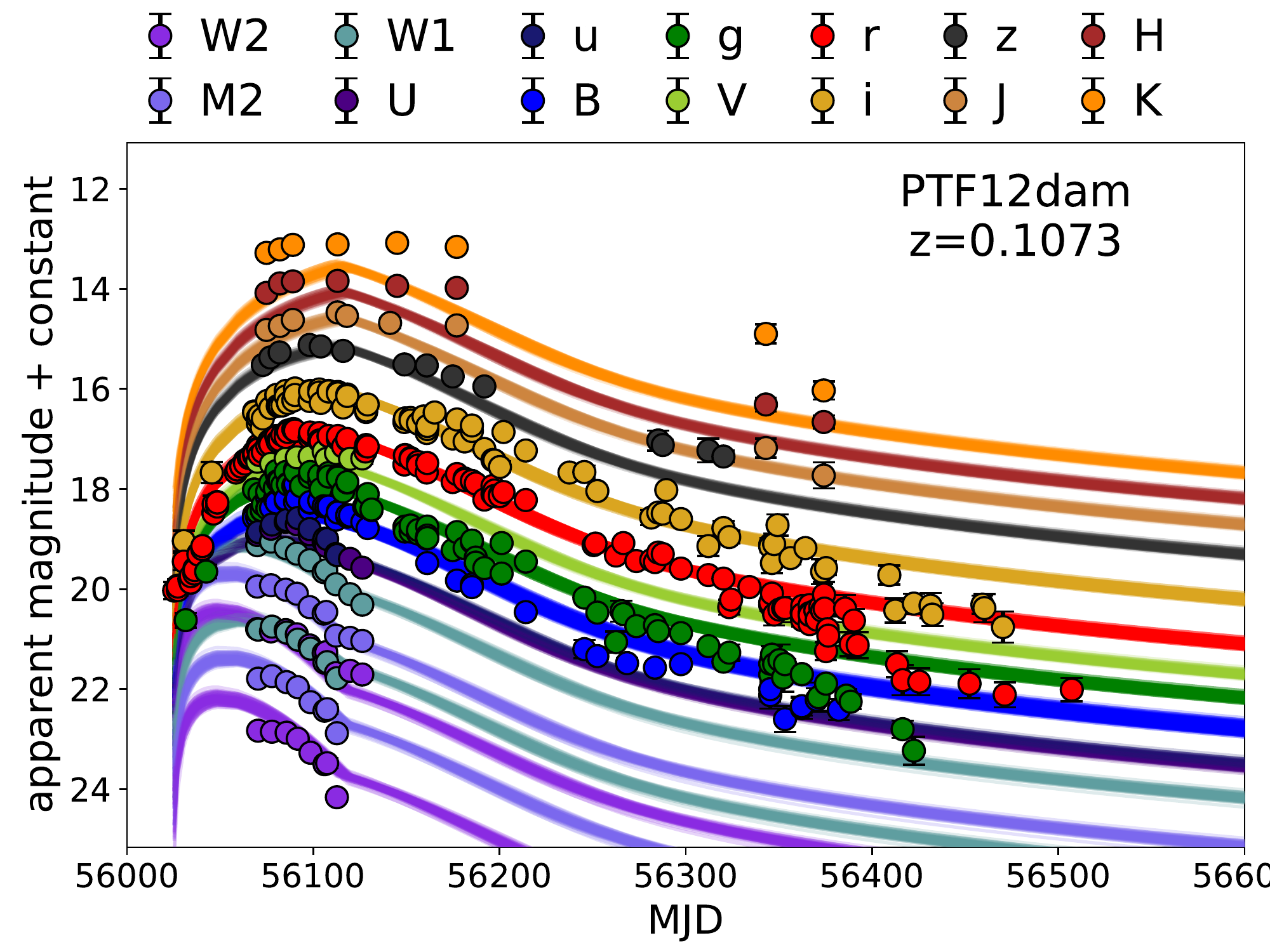}{0.33\textwidth}{}
          \fig{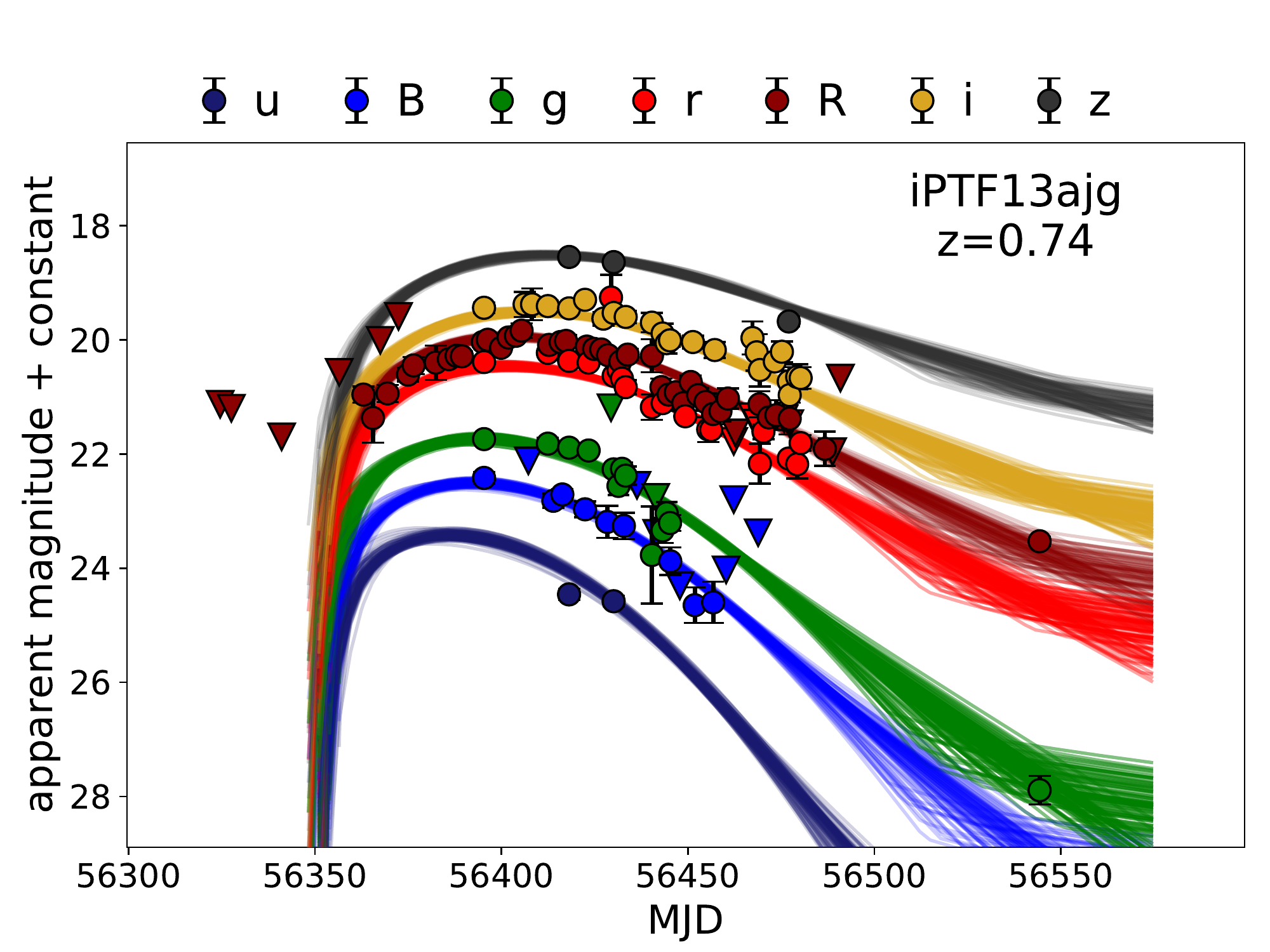}{0.33\textwidth}{}
          }
\gridline{\fig{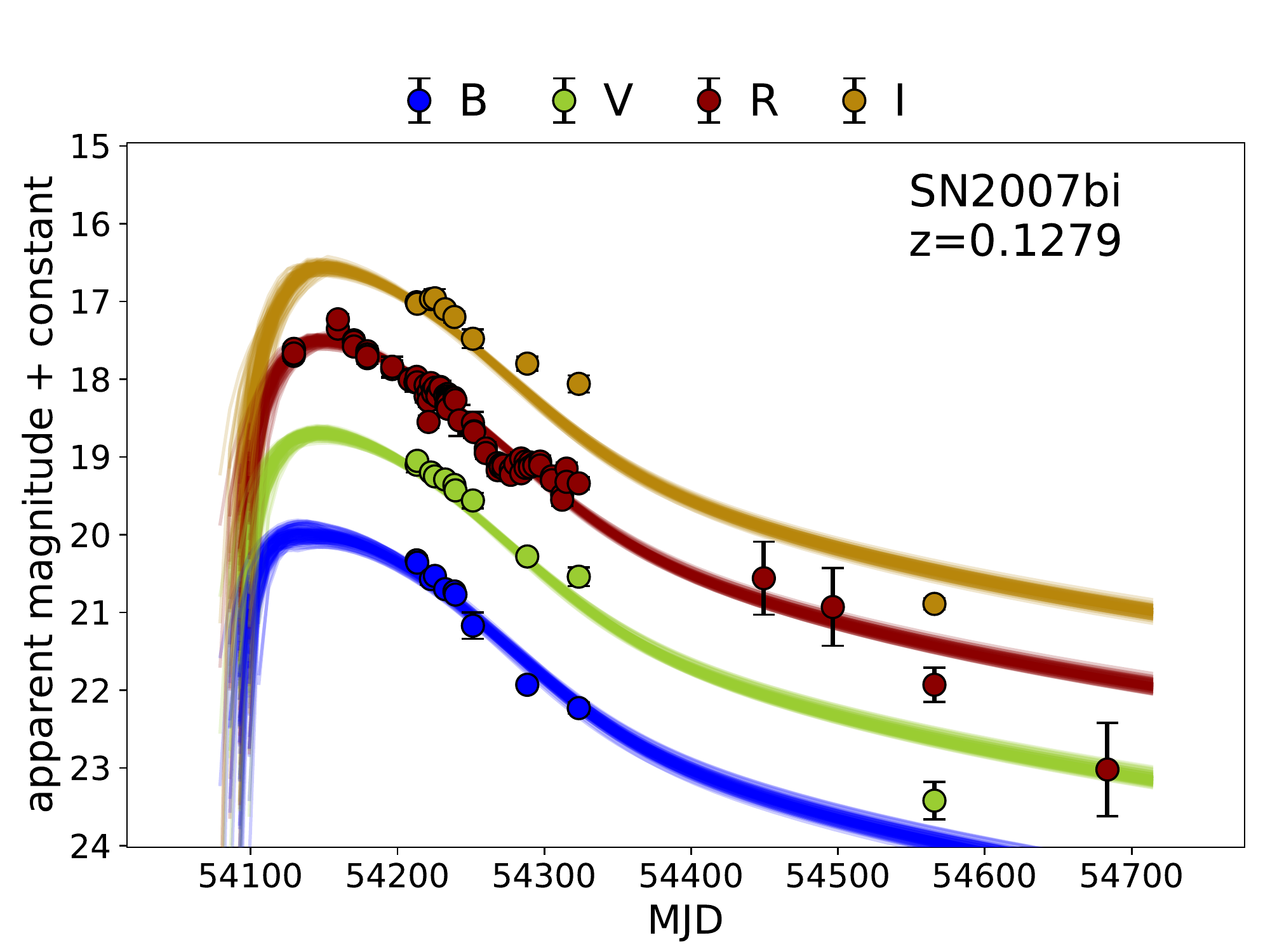}{0.33\textwidth}{}
          \fig{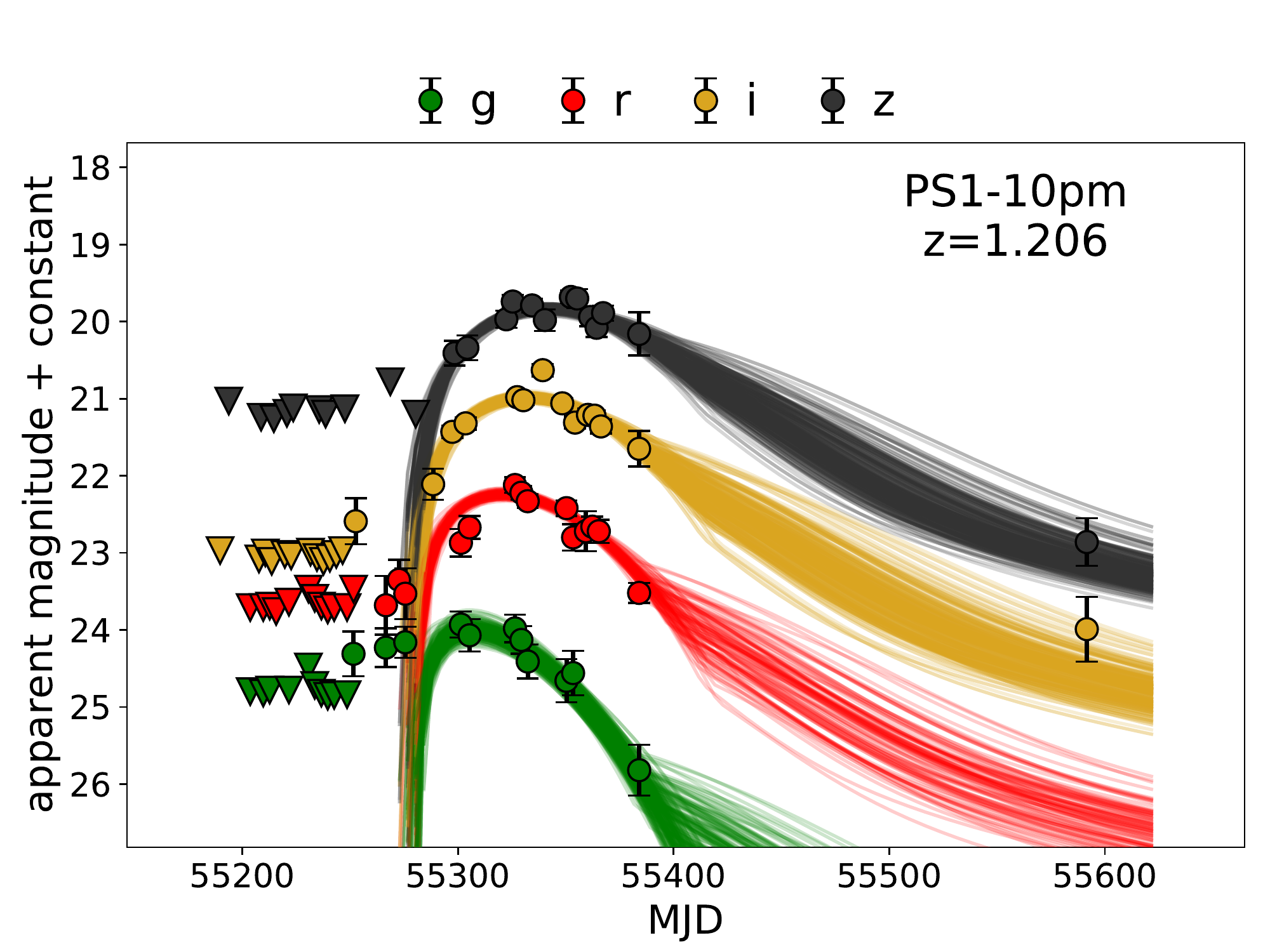}{0.33\textwidth}{}          
          \fig{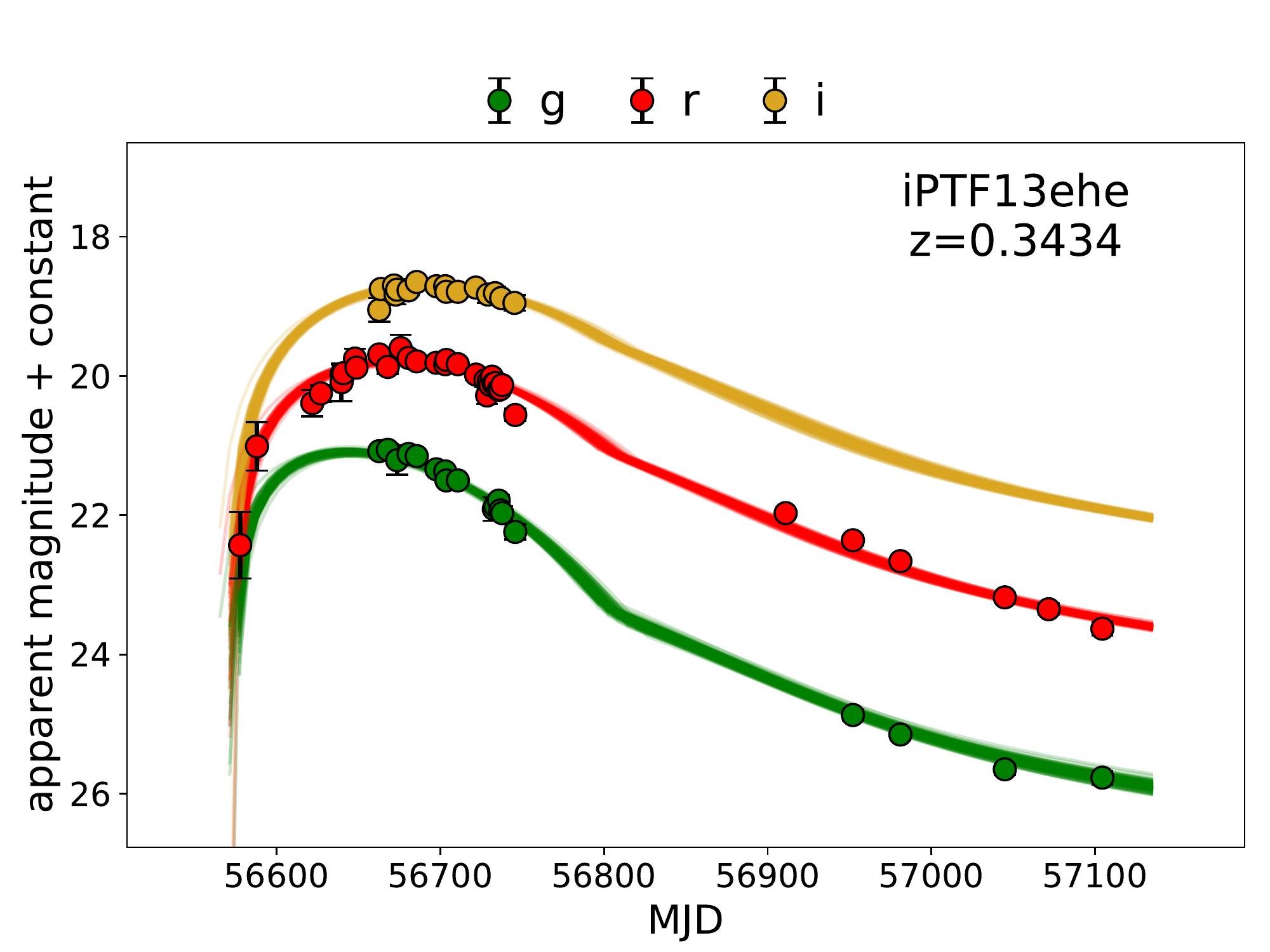}{0.33\textwidth}{}
          }
\caption{
Fallback accretion model fits to our representative SLSNe. Band offsets for display are: $uvw2+4$; $uvm2+3.5$; $uvw1+3$; $U+3$; $u+2$; $B+1.5$; $g+1$; $V+0.6$; $r+0$; $R-0.3$; $i-1$; $I-1$; $z-2$; $y-2.5$; $J-2$; $H-2.5$; $K-3$. The top two rows show SLSNe that require relatively small amount of accretion ($E_\mathrm{total}/c^2<0.01~\Msun$, red in Figure~\ref{fig:accmass}). The third row from the top shows SLSNe that require the intermediate accretion ($0.01~\Msun<E_\mathrm{total}/c^2<0.1~\Msun$, gray in Figure~\ref{fig:accmass}). The bottom row shows SLSNe that require the large accretion ($E_\mathrm{total}/c^2>0.1~\Msun$, blue in Figure~\ref{fig:accmass}).
\label{fig:lightcurves}}
\end{figure*}

\begin{figure*}[ht!]
\epsscale{1.2}
\plotone{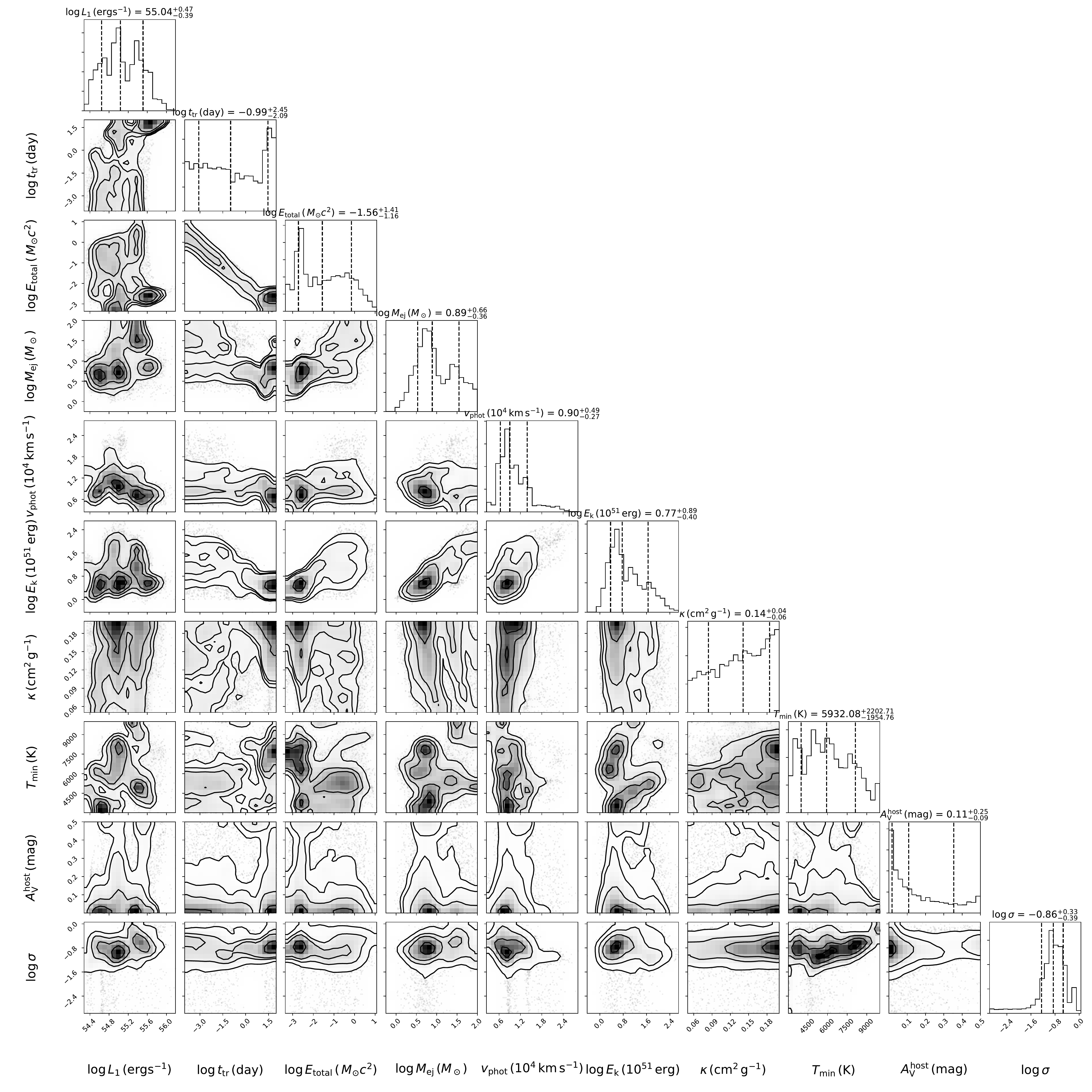}
\caption{
Joint posteriors of all model parameters acquired from our SLSN sample. Medians and $1\sigma$ ranges are shown.
\label{fig:cornerall}}
\end{figure*}

\begin{figure}[t]
\epsscale{1.2}
\plotone{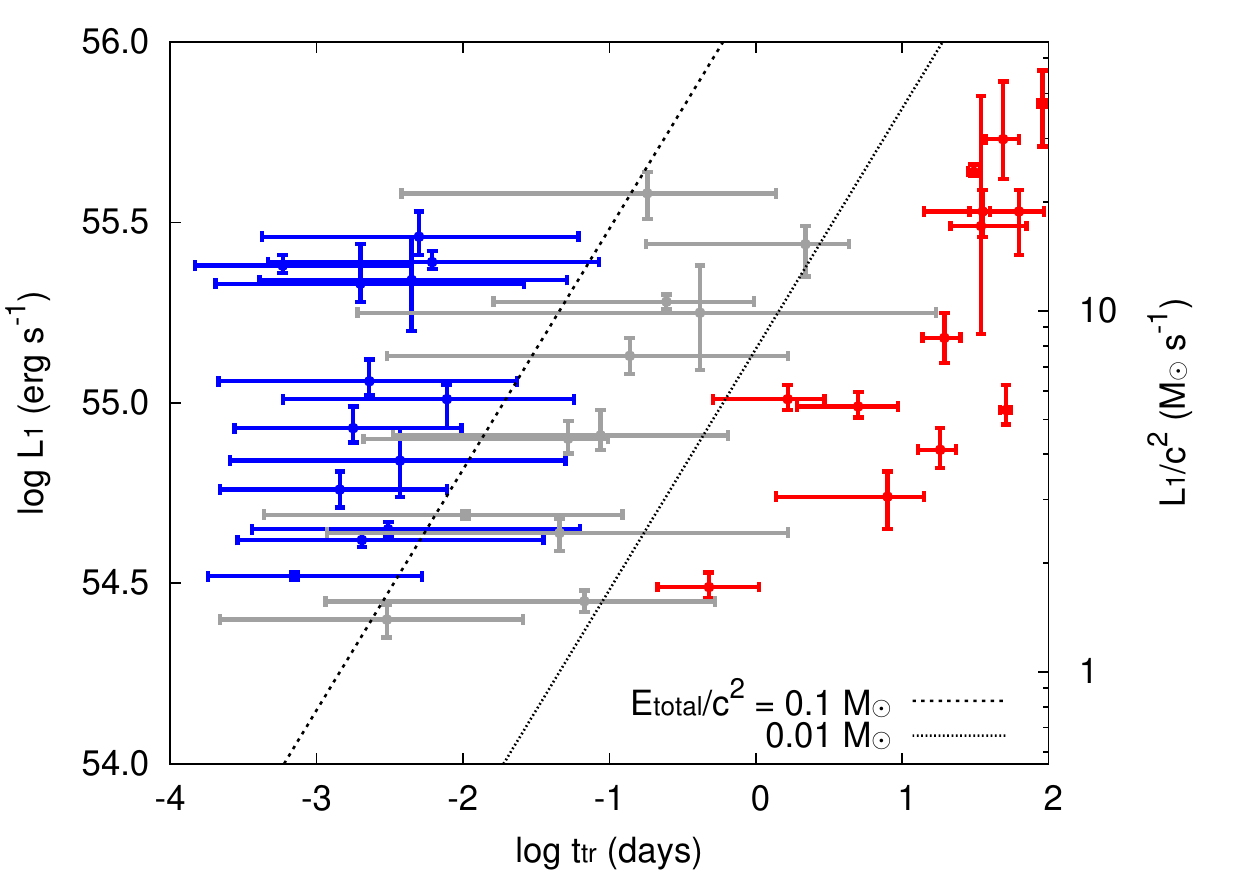}
\plotone{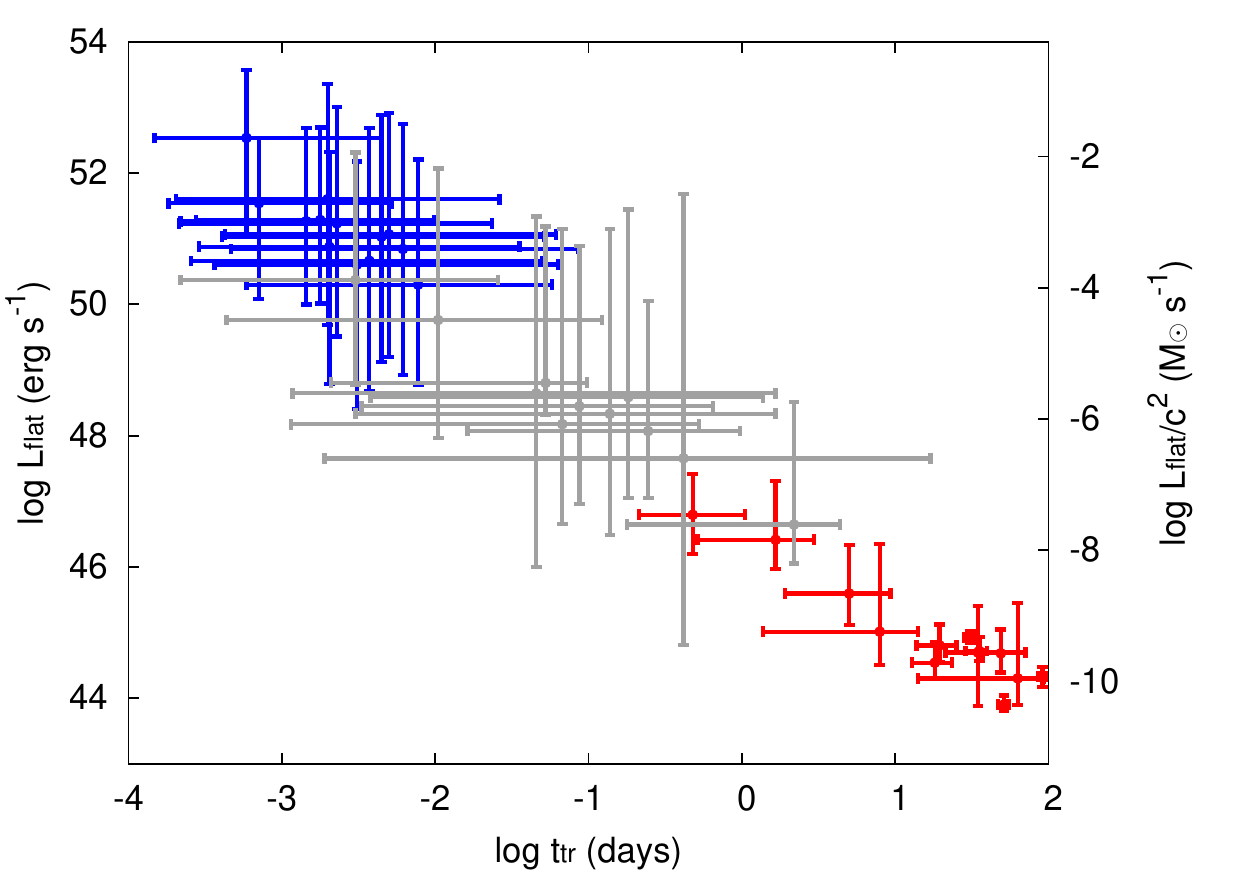}
\caption{
\textit{Top:} The fallback parameters ($t_\mathrm{tr}$ and $L_1$) estimated by \texttt{MOSFiT}. Median values and $1\sigma$ ranges are presented for every SLSN in our sample. The dotted lines show the total fallback input energy for the given set of $t_\mathrm{tr}$ and $L_1$.
\textit{Bottom:} $t_\mathrm{tr}$ and $L_\mathrm{flat}=L_1 \left(t_\mathrm{tr}/1~\mathrm{sec}\right)^{-\frac{5}{3}}$ constrained with our fitting. The SLSNe with $E_\mathrm{total}/c^2<0.01~\Msun$, $0.01~\Msun<E_\mathrm{total}/c^2<0.1~\Msun$, and $E_\mathrm{total}/c^2>0.1~\Msun$ are shown with red, gray, and blue, respectively.
\label{fig:tL}}
\end{figure}

\begin{figure}[t]
\epsscale{1.2}
\plotone{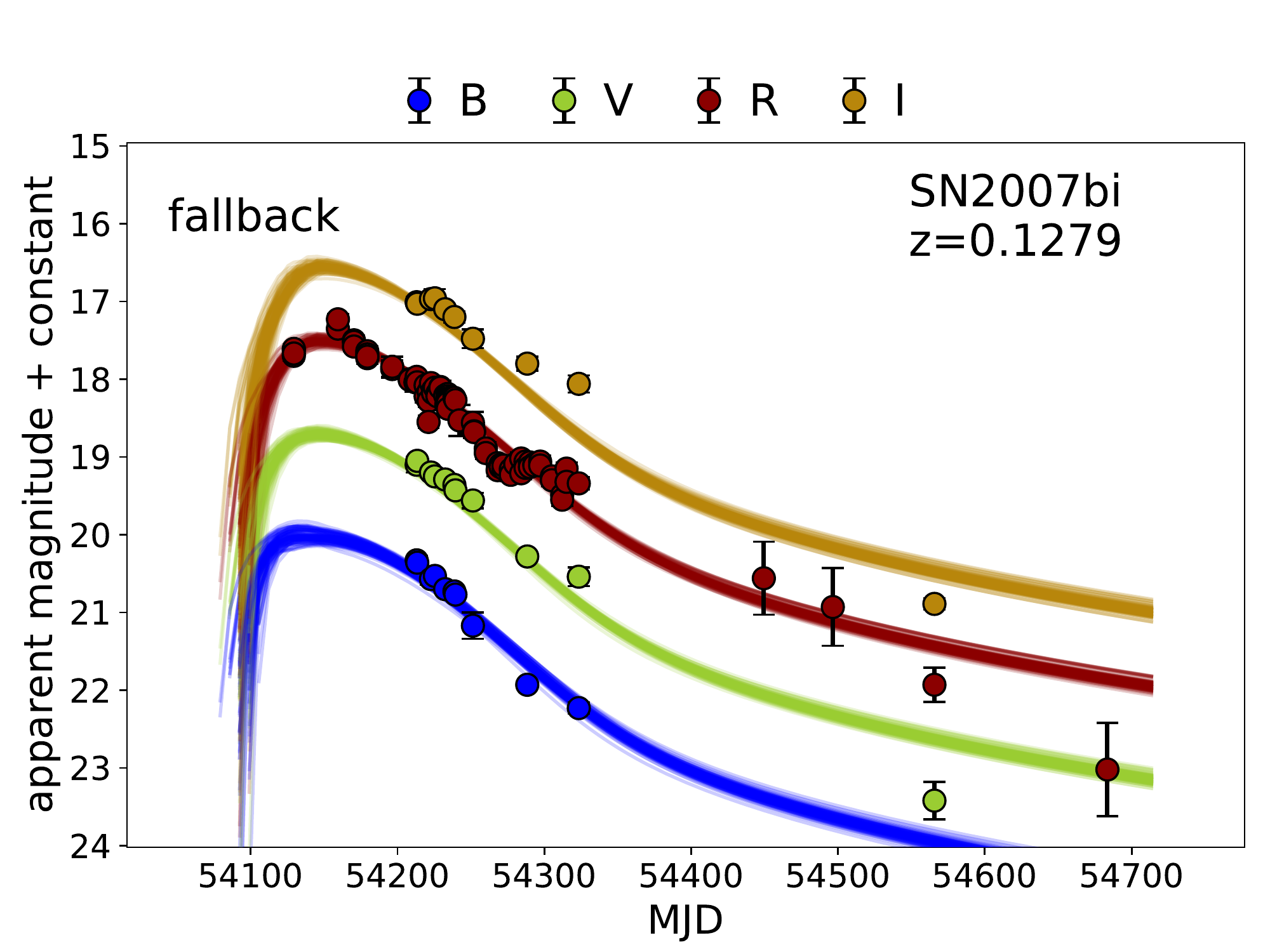}
\plotone{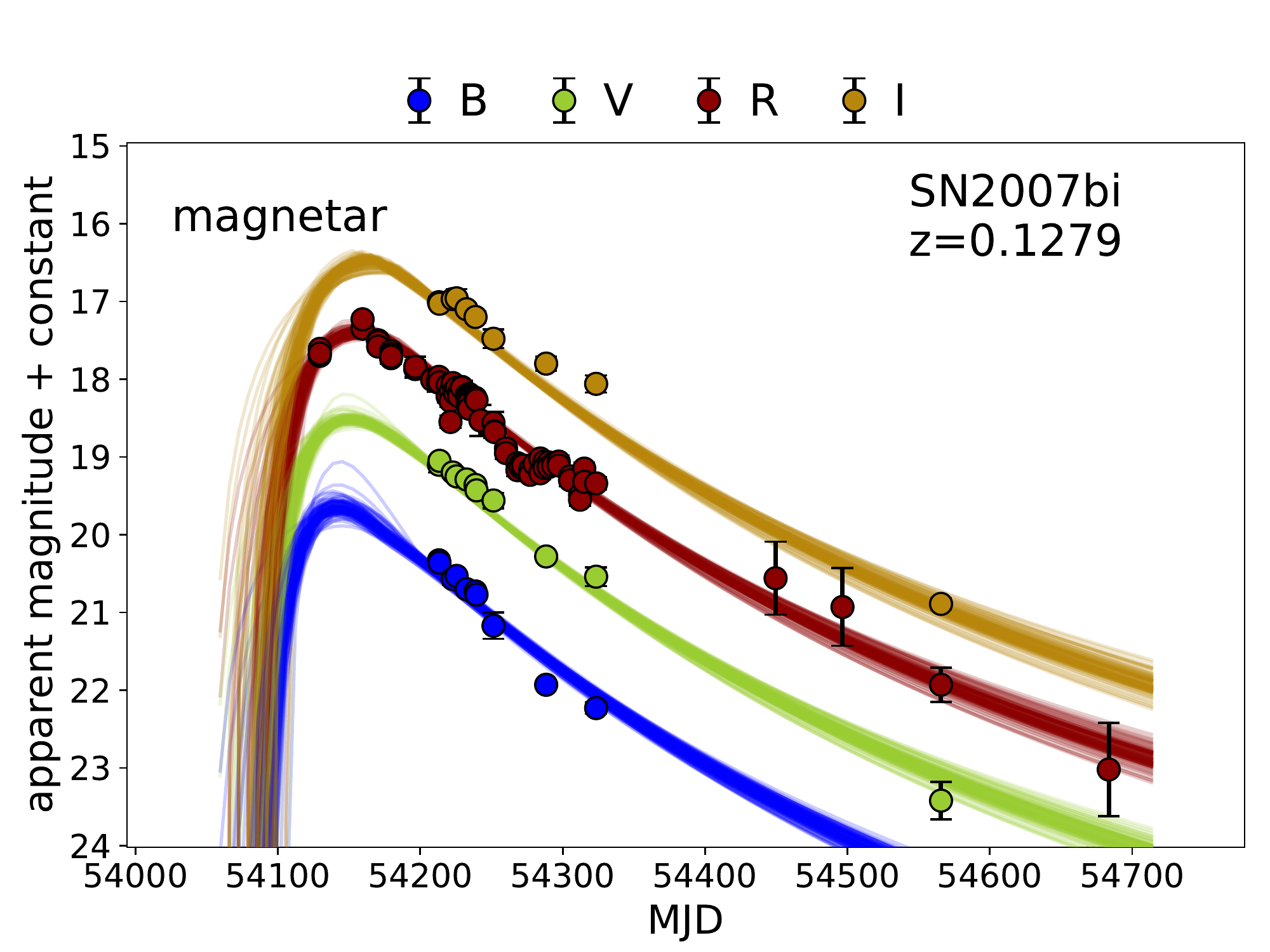}
\caption{
LC models of our fallback accretion-powered model for SN~2007bi (top) and those of the magnetar-powered model presented in \citet{nicholl2017mosmag} (bottom). Both models are similarly good at fitting the LCs. However, the magnetar model has the ejecta mass of $\Mej=3.8^{+1.4}_{-1.1}~\Msun$, while the fallback accretion model has the ejecta mass of $\Mej=67.6^{+23.6}_{-15.1}~\Msun$.
\label{fig:sn2007bifallmag}}
\end{figure}

\begin{figure}
\epsscale{1.2}
\plotone{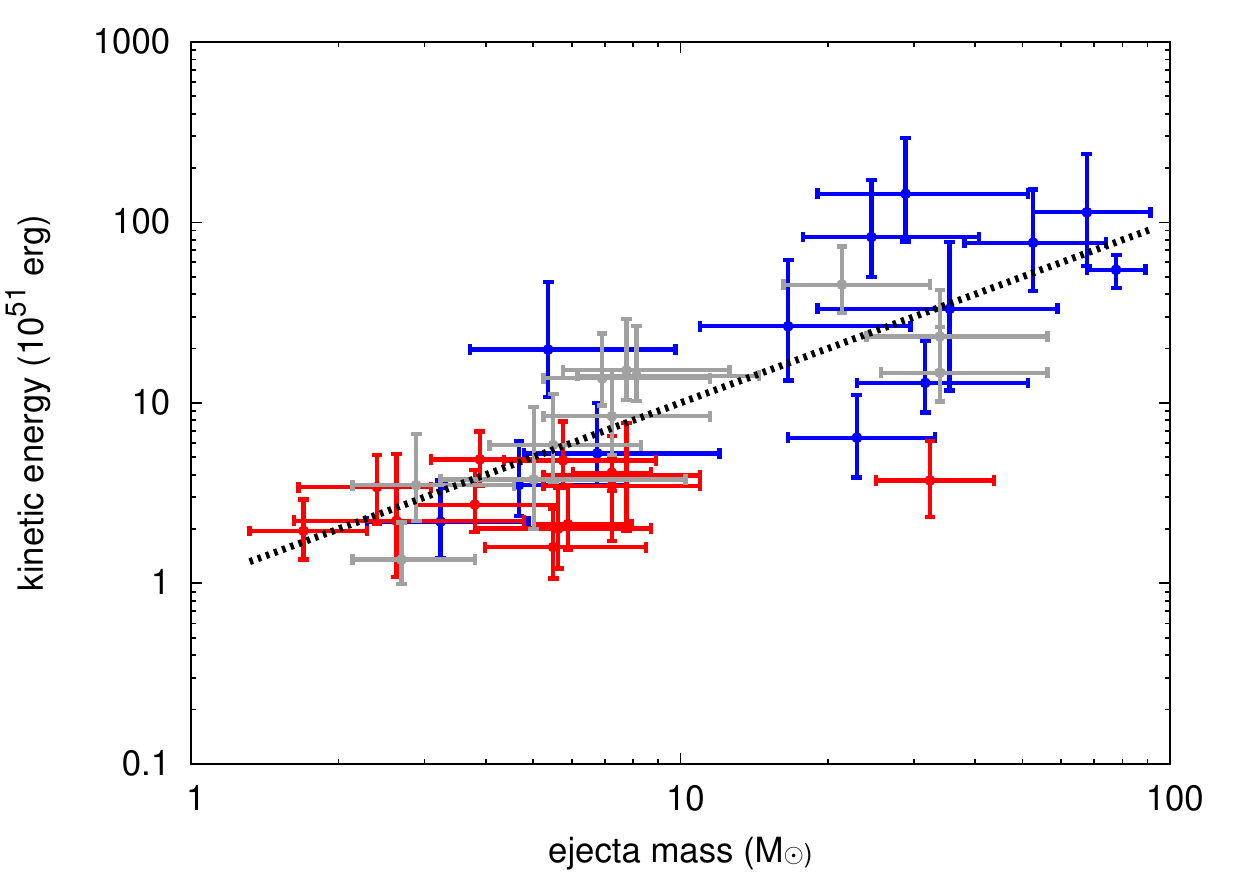}
\caption{
Ejecta mass and kinetic energy estimates of SLSNe by the fallback accretion model. Median values and $1\sigma$ ranges are presented for every SLSN in our sample. A line of $(E_\mathrm{K}/10^{51}~\mathrm{erg})/(\Mej/\Msun) = 1$, which is roughly required from spectral modeling \citep{mazzali2016slsn}, is shown. The color scheme is the same as Figure~\ref{fig:tL}.
\label{fig:kin}}
\end{figure}

\section{Results}\label{sec:results}
Figure~\ref{fig:lightcurves} shows representative results from fitting our fallback accretion model to the SLSN sample. We find that the overall quality of the LC fitting is good, showing that in principle, fallback accretion power can explain SLSN LCs. Indeed, the distribution of Watanaba-Akaike Information Criterion (WAIC, \citealt{watanabe2010waic,gelman2014waic}) does not differ much from those obtained from the magnetar-powered model presented in \citet{nicholl2017mosmag} (Table~\ref{tab:results}). Therefore, the fallback accretion model is quantitatively as good as the magnetar spin-down model in fitting SLSN LCs. We will therefore investigate the derived parameters to see whether the fallback accretion model is actually physically reasonable in the next section. 

Table~\ref{tab:results} summarizes the parameters and their standard deviations constrained by our fitting. Figure~\ref{fig:cornerall} presents the combined constraints from all SLSN fitting results. Several posteriors are bimodal in a way that is not found in the magnetar model \citep{nicholl2017mosmag}. The parameters related to the fallback accretion are constrained to be $\log (L_1/\mathrm{erg~s^{-1}}) = 55.04^{+0.47}_{-0.39}$ and $\log (t_\mathrm{tr}/\mathrm{day}) = -0.99^{+2.45}_{-2.09}$. Figure~\ref{fig:tL} presents the two fallback parameters; we find no correlation between them. We also present the first constant luminosity $L_\mathrm{flat}=L_1 \left(t_\mathrm{tr}/1~\mathrm{sec}\right)^{-\frac{5}{3}}$ in Figure~\ref{fig:tL}.
The apparent correlation between $t_\mathrm{tr}$ and $L_\mathrm{flat}$ originates from the fact that $L_1$ is well constrained with the small diversity. Because $L_1$ in $L_\mathrm{flat}=L_1 \left(t_\mathrm{tr}/1~\mathrm{sec}\right)^{-\frac{5}{3}}$ is not diverse, $L_\mathrm{flat}$ appears to have a correlation with $t_\mathrm{tr}$ as in Figure~\ref{fig:tL}.
The uncertainties in $t_\mathrm{tr}$ make $L_\mathrm{flat}$ uncertain. $L_1$ is better constrained and $L_1$ is chosen as a free parameter instead of $L_\mathrm{flat}$. From $L_1$ and $t_\mathrm{tr}$, we can derive the total input energy $E_\mathrm{total} = \int L_\mathrm{fallback}(t)dt = 2.5L_1t_\mathrm{tr}^{-2/3}$ which is also shown in Figure~\ref{fig:cornerall}. The total input energy from the fitting is $\log E_\mathrm{total}/\Msun c^2 = -1.56^{+1.41}_{-1.16}$.

The ejecta mass (\Mej) estimates are $7.8^{+27.7}_{-4.4}~\Msun$. The ejecta mass estimates are slightly larger than those found in the magnetar model ($4.8^{+8.1}_{-2.6}~\Msun$, \citealt{nicholl2017mosmag}) but, overall, the ejecta mass estimates are not much different within uncertainties in the two models. We sometimes find a large discrepancy in the two mass estimates. For example, SN~2007bi is estimated to have $\Mej=67.6^{+23.6}_{-15.1}~\Msun$ in our fallback model while the magnetar model results in $\Mej=3.8^{+1.4}_{-1.1}~\Msun$ \citep{nicholl2017mosmag}. The results of the LC fitting are not much different from each other in the two models (Figure~\ref{fig:sn2007bifallmag}). We found that the large mass discrepancy tends to appear when the spin-down timescale in the magnetar model is larger than the diffusion timescale. In such a case, the rise time of the magnetar model is strongly affected by the spin-down timescale, not by the diffusion timescale that strongly affects the ejecta mass estimate \citep[e.g.,][]{kasen2010mag}.

To have an idea of the kinetic energy of the ejecta, we assume $E_\mathrm{K}=\Mej v_\mathrm{phot}^2/2$. Because $v_\mathrm{phot}$ is just a photospheric velocity, $E_\mathrm{K}$ is unlikely to be the true kinetic energy of SN ejecta, but it provides a rough approximation \citep[e.g.,][]{arnett1982}. Figure~\ref{fig:kin} shows the relation between the ejecta mass and the kinetic energy. We find that SLSNe requiring higher ejecta masses tend to have higher kinetic energies. Spectral modeling of SLSNe indicates that $(E_\mathrm{K}/10^{51}~\mathrm{erg})/(\Mej/\Msun) \sim 1$ in SLSNe \citep{mazzali2016slsn,liu2017slsnspec,howell2013slsn} and a similar ratio was found from light curve modelling using a magnetar engine \citep{nicholl2017mosmag}.
We find our results also follow this trend. In the fallback accretion scenario, the \emph{initial} explosion energy should be relatively small, in order to achieve significant fallback. However, the ejecta that do escape should gain additional energy from the accretion power after the explosion, such that it eventually reaches $(E_\mathrm{K}/10^{51}~\mathrm{erg})/(\Mej/\Msun) \sim 1$.

\begin{figure}
\epsscale{1.2}
\plotone{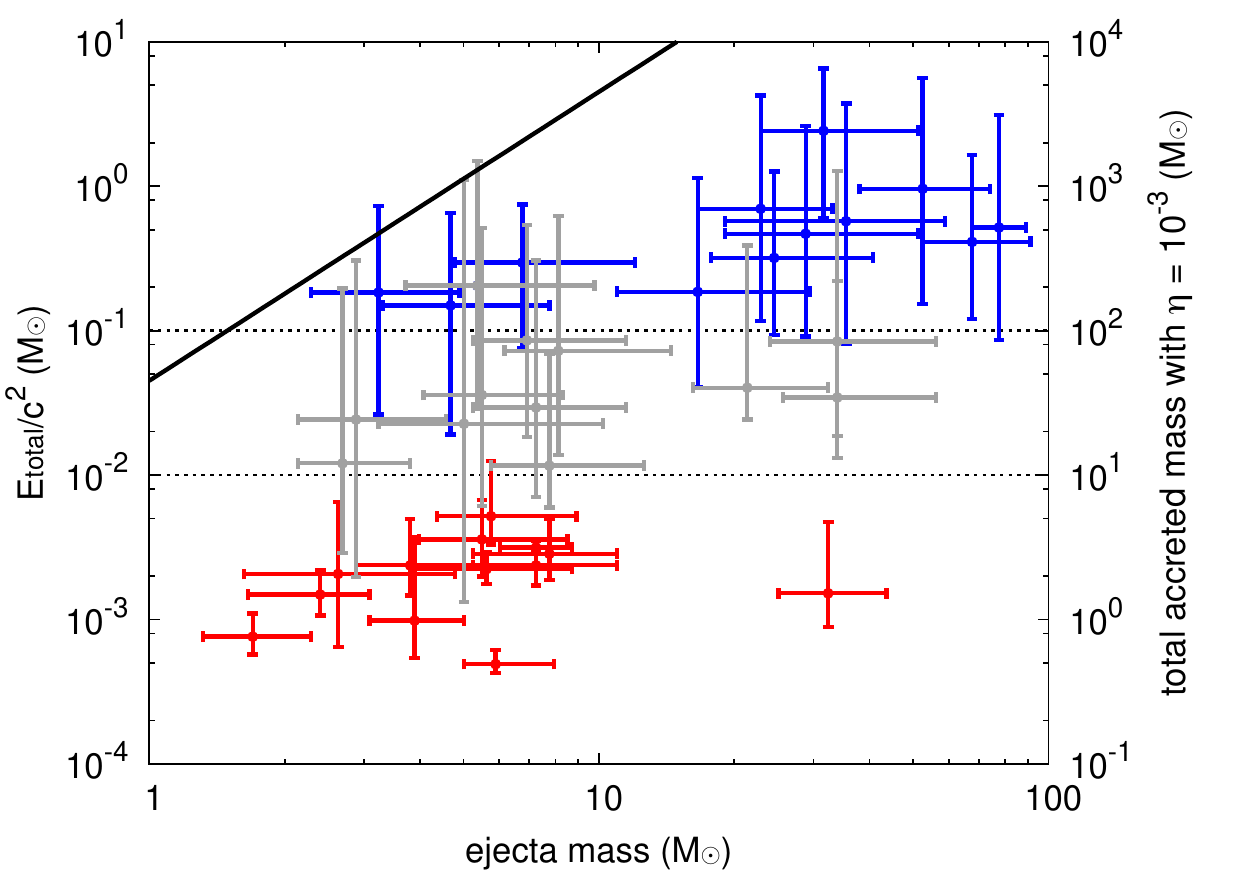}
\caption{
The estimated total required accretion mass and the ejecta mass. Median values and $1\sigma$ ranges are presented for every SLSN in our sample. $E_\mathrm{total}/c^2$ corresponds to total amount of mass required to convert to central input energy. The right vertical axis shows the required total accretion mass assuming the conversion efficiency of $\eta = 10^{-3}$. The color scheme is the same as Figure~\ref{fig:tL}. The solid line indicates the total kinetic energy carried by the disk outflow having the same mass as the ejecta mass, assuming the outflow velocity of $0.3c$.
\label{fig:accmass}}
\end{figure}

\begin{figure}
\epsscale{1.2}
\plotone{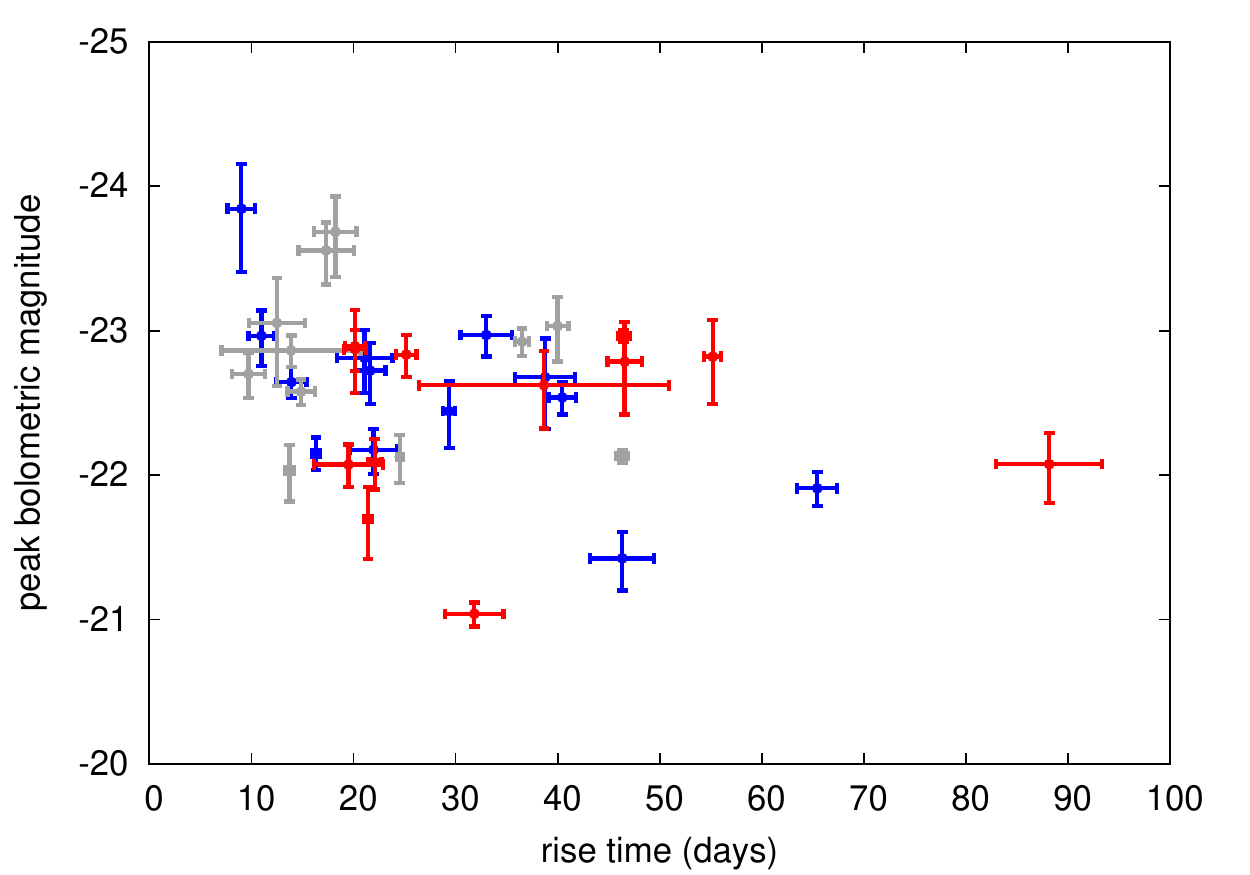}
\caption{
Rise time of the bolometric LC and the peak bolometric magnitude of the SLSNe in our sample. The color scheme is the same as Figure~\ref{fig:tL}.
\label{fig:risemag}}
\end{figure}

\begin{figure}
\epsscale{1.2}
\plotone{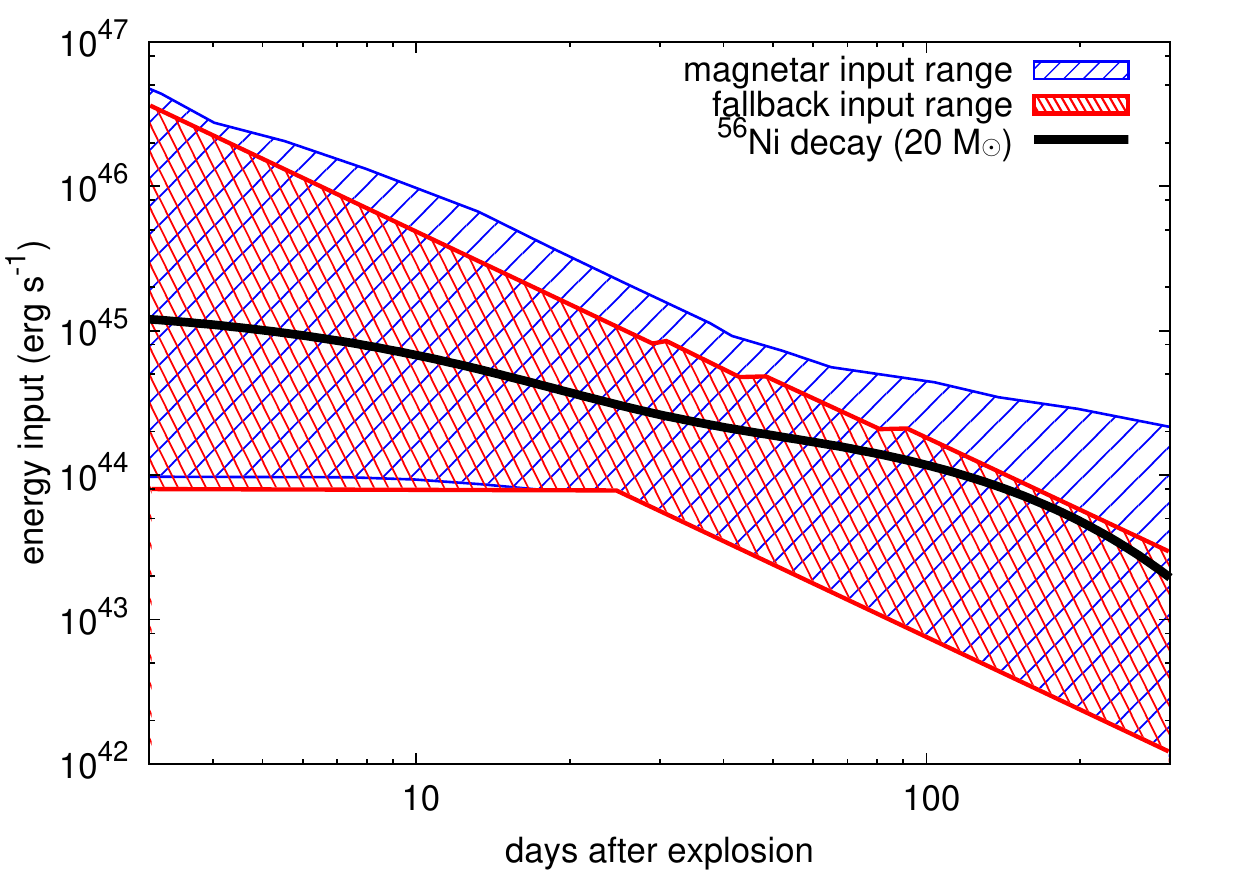}
\caption{
The regions of the central energy inputs required to fit the SLSN LCs. The region for the fallback accretion energy input is obtained in this work and the region region occupied by the magnetar central energy input is from \citet{nicholl2017mosmag}. The decay energy of 20~\Msun\ of \Ni\ is shown for comparison \citep{nadyozhin1994nico}.
\label{fig:magcomp}}
\end{figure}

\section{Discussion}\label{sec:discussion}
The total central energy input required to power SLSNe ($E_\mathrm{total}$) is summarized in Figure~\ref{fig:accmass}. We find that $E_\mathrm{total}/ c^2\sim 10^{-3} - 1~\Msun$ must be liberated in the accretion process. The essential question then is how much total mass needs to be accreted to the central compact remnant in order to produce this amount of energy. If we set $\eta$ as the conversion efficiency from accretion, the required accretion mass becomes $E_\mathrm{total}/\eta c^2$. \citet{dexter2013kasen} estimate that the typical conversion efficiency from the fallback accretion to the large-scale outflow is $\eta\sim 10^{-3}$ by using the disk accretion models. Using this efficiency with our derived energies, we find that the total mass accreted must be $\sim 1 - 1000~\Msun$ (Figure~\ref{fig:accmass}). Accreting $\sim 1  - 10~\Msun$ of material from the carbon and oxygen core might be possible if the core is massive enough \citep[e.g.,][]{aguilera-dena2018slsngrb}. However, having carbon and oxygen cores of $\sim 100~\Msun$ is very challenging even in low-metallicity environments \citep[e.g.,][]{yoshida2014massiveco}. Therefore, most SLSNe likely require too great an accretion mass to be explained by fallback accretion.

Some region of parameter space likely remains available for the fallback model as the true efficiency $\eta$ is quite uncertain. \citet{dexter2013kasen} estimated $\eta \sim 10^{-3}$ with a plausible parameter set, assuming a large-scale disk outflow. Alternatively, if we assume that the major source of the outflow is a jet launched at the inner edge of the accretion disk, the efficiency might be as high as $\eta\sim 0.1$ \citep[e.g.,][]{mckinney2005jeteff,kumar2008fallbackgrb,gilkis2016jetfeedback}. In this extreme, most SLSNe would require a more reasonable accreted mass of less than 10~\Msun.

Even with an inefficient conversion efficiency, there are some SLSNe for which we need only a modest accreted mass to account for their LCs, and so could still be powered by fallback accretion. Those SLSNe that only require the accretion of $E_\mathrm{total}/c^2<0.01~\Msun$, indicated with red in Figures~\ref{fig:tL}, \ref{fig:kin}, \ref{fig:accmass}, and \ref{fig:risemag}, tend to have low kinetic energy. However, both intermediate SLSNe ($0.01~\Msun<E_\mathrm{total}/c^2<0.1~\Msun$, gray in the figures) and SLSNe requiring large accretion mass ($E_\mathrm{total}/c^2>0.1~\Msun$, blue in the figures) can sometimes have low kinetic energy too, so kinetic energy alone is not a robust diagnostic of fallback candidates.
Figure~\ref{fig:risemag} shows the rise time in the bolometric LC and the peak bolometric magnitude of the SLSNe in our sample. They are obtained from our fitting results. The SLSNe with lower accretion masses do not have clear distinction from the SLSNe with the higher accretion masses. The peak luminosity is found not to have a strong relation with the accreted mass, although the brightest events seem to require high accretion and it might be hard for the fallback model to explain them with physically plausible parameters (Figure~\ref{fig:risemag}).

The transitional time $t_\mathrm{tr}$ could be physically related to the freefall accretion timescale which is roughly $\left(G \rho\right)^{-1/2}$. $t_\mathrm{tr}$ ranges from $\sim 0.1~\mathrm{days}$ to $\sim 100~\mathrm{days}$ among the SLSNe that require the smallest amount of accretion (red in the figures) and, therefore, are the plausible fallback accretion-powered SLSN candidates. $t_\mathrm{tr}\sim 0.1~\mathrm{days}$ corresponds to the main-sequence-like density $\rho\sim 0.1~\mathrm{g~cm^{-3}}$ and $t_\mathrm{tr}\sim 100~\mathrm{days}$ corresponds to the giant-like density $\rho\sim 10^{-6}~\mathrm{g~cm^{-3}}$. $t_\mathrm{tr}$ may imply such a progenitor but it may also affected by the explosion dynamics such as reverse shocks \citep[e.g.,][]{zhang2008acc,dexter2013kasen}.

The necessity of a large fallback accretion rate with $L_1\sim 10^{55}~\mathrm{erg~s^{-1}}$ in fitting the SLSN LCs can be understood in a simple way. SLSNe have rise times of $20-100~\mathrm{days}$ and peak luminosities of the order of $10^{44}~\mathrm{erg~s^{-1}}$ \citep{nicholl2015slsndiv,lunnan2018slsnps1,decia2017slsnptf}. A simple analytic model finds that the peak luminosity powered by a central heating source matches the energy input at the LC peak (the so-called Arnett law, \citealt{arnett1979thelaw}). Figure~\ref{fig:magcomp} shows a comparison between the range of the fallback accretion power and the range of the magnetar spin-down power required to fit SLSNe \citep{nicholl2017mosmag}. We can see in the figure that the two central energy inputs found to explain SLSN LCs occupy almost the same luminosity range especially at $20-100~\mathrm{days}$ when the LCs reach their peak. Following the Arnett law, we can say that energy input as in Figure~\ref{fig:magcomp} is generally required to power SLSNe with any central engines. The Arnett law is a simplified formalism, but the necessity of long-sustained high accretion rates in the fallback accretion model, as estimated by our Bayesian approach, is unavoidable.

In this work, we treat \Mej\ as a constant parameter in the fitting procedure. However, \Mej\ can change as a function of time in the actual fallback accretion powered SNe. One reason is that the disk outflow is the origin of the thermal energy and the flown material eventually becomes a part of the SN ejecta. In Figure~\ref{fig:accmass}, we draw the solid line that shows the total energy that is released with the corresponding ejecta mass assuming the typical disk outflow velocity of $0.3c$. The SLSNe on this line should have the outflow as massive as the ejecta mass. A large fraction of SLSNe locate well below the line and the change in the ejecta mass is expected to be rather small. Even if the ejecta mass is increased by the disk wind, the necessity of the large total ejecta mass to account for the large diffusion time would not change. Another reason is that the power source of our model is the mass accretion and the ejecta mass should be reduced with the accretion. We often find that the required accreted mass is larger than the ejecta mass with the assumption of $\eta = 10^{-3}$. The actual mass reduced from the ejecta strongly depends on the uncertain $\eta$. Regardless of the uncertainty in $\eta$, our conclusion that the required accretion mass becomes too large would remain in the cases where the required accretion mass starts to dominate the ejecta mass.

Although we find that the fallback model is statistically as good as the magnetar model in fitting the SLSN LCs, the two models may have significant differences in their expected spectroscopic properties. The fallback accretion model often requires the accretion of $\gtrsim1~\Msun$  to the central compact remnant. This means that most of the inner core of the progenitor would be accreted. Therefore a SLSN that shows a lack of heavy element signatures in the spectrum would be a possible fallback candidate. So far, SLSNe with late-time spectra do not show a deficiency of iron-group elements relative to other SNe \citep{gal-yam2009sn2007bi,nicholl2016sn2015bnneb,jerkstrand2017slsnneb}, and SLSNe even likely to produce more iron than other SNe \citep{nicholl2018slsnnebular}.
Another possible smoking gun is an off-axis afterglow from the jet that is likely produced when the accretion disk is highly super-Eddington. However, no such afterglows have been found from SLSNe and no afterglows have shown SN counterparts that have SLSNe luminosities, although one afterglow from a ultra-long gamma-ray burst is found to have a bright SN component \citep{greiner2015ulgrbslsn}.

\section{Conclusions}\label{sec:conclusions}
We have systematically investigated the fallback accretion central engine model for hydrogen-poor SLSNe. By using \texttt{MOSFiT}, we have fitted the multi-band LCs of 37 SLSNe, finding that the model provides satisfactory fits to the full ensemble of SLSN LCs. The quality of the LC fits are quantitatively as good as a similar model powered by magnetar spin-down, previously investigated with the same approach \citep{nicholl2017mosmag}.

However, we have found that the total energy input from the fallback accretion that needs to be provided to power the SLSNe is $0.002-0.7~\Msun c^2$ (Figure~\ref{fig:accmass}). Assuming a realistic conversion efficiency from the fallback accretion disk to the large-scale outflow ($\sim 10^{-3}$), the total mass that must be accreted is $2-700~\Msun$. Therefore, this model often requires too much accretion to be achieved by massive stars. Thus we conclude that fallback is unlikely to power the majority of SLSNe.

The conversion efficiency is uncertain, and if it could reach $\sim 0.1$ in some cases, the required accretion mass might be limited to $0.02-7~\Msun$. However, realistic simulations are needed to see if such a high value could be attained in practice.

Regardless of the uncertain efficiency, there are some SLSNe for which the required accretion is low enough that it could be compatible with massive star collapse. We find that they are difficult to be distinguished by LCs (Figure~\ref{fig:risemag}). The lack of the heavy elements in the nebular phase spectra or afterglow observations may identify the fallback powered SNe, although the recent studies show that SLSNe likely produce more iron than ordinary SNe \citep{nicholl2018slsnnebular}.

\acknowledgments
We thank the referee for the constructive comments that improved this work.
TJM thanks Tomohisa Kawashima and David Aguilera-Dena for very useful discussion.
TJM and MN acknowledge the support from the Munich Institute for Astro- and Particle Physics (MIAPP) of the DFG cluster of excellence "Origin and Structure of the Universe."
TJM is supported by the Grants-in-Aid for Scientific Research of the Japan Society for the Promotion of Science (16H07413, 17H02864, 18K13585).
Numerical computations were in part carried out on computers at Center for Computational Astrophysics, National Astronomical Observatory of Japan.

%

\vspace{5mm}


\software{MOSFiT \citep{guillochon2017mosfit}
          }



\begin{longrotatetable}
\begin{deluxetable*}{lrrrrrrrrrrr}
\tablecaption{Medians and $1\sigma$ bounds for the fitted parameters. \label{tab:results}}
\tablewidth{700pt}
\tabletypesize{\tiny}
\tablehead{
\colhead{Name} & \colhead{$\log L_1$} & 
\colhead{$\log t_\mathrm{tr}$} & \colhead{$\log E_\mathrm{total}$} & \colhead{$\Mej$} & 
\colhead{$v_\mathrm{phot}$} & \colhead{$E_\mathrm{K}$} & 
\colhead{$\kappa_\ej$} & 
\colhead{$T_\mathrm{f}$} & \colhead{$A_V$} & \colhead{$\sigma$} & \colhead{WAIC} \\ 
\colhead{} & \colhead{($\mathrm{erg~s^{-1}}$)} & \colhead{(day)} & \colhead{($\Msun c^2$)} & \colhead{(\Msun)} & 
\colhead{(1000~\kmps)} & \colhead{($10^{51}$~erg)} & \colhead{($\mathrm{cm^2~g^{-1}}$)} &
\colhead{(1000 K)} & \colhead{(mag)} & \colhead{} & \colhead{}
} 
\startdata
SN~2005ap   & $54.84^{+0.09}_{-0.10}$ & $-2.43^{+1.13}_{-1.16}$ & $ -0.69^{+  0.86}_{-  0.85}$ & $  5.37^{+  4.40}_{-  1.65}$ & $ 19.17^{+  2.73}_{-  2.14}$ & $   19.73^{+   27.14}_{-    8.96}$ & $0.12^{+0.05}_{-0.05}$ & $  4.84^{+  0.62}_{-  0.75}$ & $  0.21^{+  0.18}_{-  0.14}$ & $  0.01^{+  0.01}_{-  0.00}$ & $   56.22$ \\
SN~2006oz   & $55.49^{+0.36}_{-0.30}$ & $1.54 ^{+0.31}_{-0.21}$ & $ -2.69^{+  0.50}_{-  0.51}$ & $  2.63^{+  2.16}_{-  1.01}$ & $  9.19^{+  1.25}_{-  1.01}$ & $    2.22^{+    2.99}_{-    1.14}$ & $0.14^{+0.05}_{-0.05}$ & $  5.92^{+  1.02}_{-  1.11}$ & $  0.16^{+  0.20}_{-  0.11}$ & $  0.01^{+  0.01}_{-  0.00}$ & $   75.30$ \\
SN~2007bi   & $54.93^{+0.06}_{-0.04}$ & $-2.75^{+0.74}_{-0.81}$ & $ -0.39^{+  0.60}_{-  0.53}$ & $ 67.61^{+ 23.59}_{- 15.13}$ & $ 12.96^{+  3.22}_{-  2.51}$ & $  113.53^{+  125.14}_{-   56.23}$ & $0.16^{+0.03}_{-0.04}$ & $  8.47^{+  0.37}_{-  0.29}$ & $  0.10^{+  0.12}_{-  0.08}$ & $  0.14^{+  0.01}_{-  0.01}$ & $  162.35$ \\
SN~2010gx   & $54.69^{+0.01}_{-0.01}$ & $-1.98^{+1.07}_{-1.38}$ & $ -1.14^{+  0.93}_{-  0.72}$ & $  8.13^{+  6.33}_{-  1.96}$ & $ 13.19^{+  0.38}_{-  0.28}$ & $   14.14^{+   12.46}_{-    3.87}$ & $0.14^{+0.05}_{-0.06}$ & $  3.62^{+  0.24}_{-  0.22}$ & $  0.01^{+  0.02}_{-  0.01}$ & $  0.14^{+  0.01}_{-  0.01}$ & $  247.49$ \\
SN~2011ke   & $54.52^{+0.01}_{-0.01}$ & $-3.15^{+0.87}_{-0.59}$ & $ -0.53^{+  0.40}_{-  0.59}$ & $  6.76^{+  5.26}_{-  1.97}$ & $  8.80^{+  0.30}_{-  0.27}$ & $    5.24^{+    4.72}_{-    1.76}$ & $0.12^{+0.05}_{-0.05}$ & $  3.89^{+  0.16}_{-  0.18}$ & $  0.01^{+  0.01}_{-  0.01}$ & $  0.31^{+  0.01}_{-  0.01}$ & $   95.59$ \\
SN~2011kf   & $54.76^{+0.05}_{-0.05}$ & $-2.84^{+0.73}_{-0.82}$ & $ -0.50^{+  0.60}_{-  0.54}$ & $ 24.55^{+ 16.19}_{-  6.76}$ & $ 18.39^{+  2.12}_{-  1.64}$ & $   83.00^{+   88.38}_{-   33.10}$ & $0.13^{+0.05}_{-0.05}$ & $  5.48^{+  0.13}_{-  0.11}$ & $  0.03^{+  0.05}_{-  0.03}$ & $  0.07^{+  0.02}_{-  0.02}$ & $   65.76$ \\
SN~2012il   & $54.62^{+0.03}_{-0.02}$ & $-2.69^{+1.24}_{-0.85}$ & $ -0.74^{+  0.60}_{-  0.85}$ & $  3.24^{+  1.66}_{-  0.95}$ & $  8.23^{+  0.48}_{-  0.49}$ & $    2.19^{+    1.53}_{-    0.82}$ & $0.11^{+0.04}_{-0.04}$ & $  6.02^{+  0.20}_{-  0.15}$ & $  0.05^{+  0.06}_{-  0.03}$ & $  0.10^{+  0.02}_{-  0.01}$ & $   60.57$ \\
SN~2013dg   & $54.65^{+0.02}_{-0.02}$ & $-2.51^{+1.31}_{-0.93}$ & $ -0.83^{+  0.64}_{-  0.89}$ & $  4.68^{+  3.09}_{-  1.37}$ & $  8.65^{+  0.22}_{-  0.22}$ & $    3.50^{+    2.62}_{-    1.15}$ & $0.12^{+0.05}_{-0.05}$ & $  3.02^{+  0.04}_{-  0.02}$ & $  0.06^{+  0.07}_{-  0.04}$ & $  0.01^{+  0.02}_{-  0.01}$ & $  128.50$ \\
SN~2013hy   & $54.98^{+0.07}_{-0.04}$ & $1.71 ^{+0.03}_{-0.04}$ & $ -3.31^{+  0.10}_{-  0.06}$ & $  5.89^{+  2.05}_{-  0.88}$ & $  6.00^{+  0.63}_{-  0.48}$ & $    2.12^{+    1.37}_{-    0.60}$ & $0.16^{+0.03}_{-0.05}$ & $  7.65^{+  0.35}_{-  0.32}$ & $  0.08^{+  0.13}_{-  0.06}$ & $  0.12^{+  0.01}_{-  0.02}$ & $  124.29$ \\
SN~2015bn   & $55.38^{+0.03}_{-0.02}$ & $-3.23^{+0.88}_{-0.60}$ & $  0.38^{+  0.43}_{-  0.61}$ & $ 31.62^{+ 19.66}_{-  8.71}$ & $  6.38^{+  0.19}_{-  0.17}$ & $   12.87^{+    9.28}_{-    4.04}$ & $0.13^{+0.04}_{-0.05}$ & $  8.89^{+  0.25}_{-  0.19}$ & $  0.04^{+  0.05}_{-  0.03}$ & $  0.30^{+  0.01}_{-  0.01}$ & $  313.28$ \\
PTF09atu    & $55.58^{+0.06}_{-0.07}$ & $-0.74^{+0.88}_{-1.68}$ & $ -1.08^{+  1.18}_{-  0.66}$ & $ 33.88^{+ 22.35}_{-  9.90}$ & $  8.30^{+  0.36}_{-  0.41}$ & $   23.35^{+   18.79}_{-    8.40}$ & $0.12^{+0.05}_{-0.04}$ & $  5.10^{+  0.39}_{-  0.65}$ & $  0.38^{+  0.09}_{-  0.13}$ & $  0.11^{+  0.01}_{-  0.01}$ & $  157.03$ \\
PTF09cnd    & $55.64^{+0.02}_{-0.01}$ & $1.49 ^{+0.04}_{-0.04}$ & $ -2.50^{+  0.05}_{-  0.04}$ & $  7.24^{+  1.47}_{-  1.22}$ & $  7.52^{+  0.14}_{-  0.18}$ & $    4.10^{+    1.02}_{-    0.85}$ & $0.16^{+0.03}_{-0.03}$ & $  3.97^{+  0.26}_{-  0.32}$ & $  0.01^{+  0.01}_{-  0.00}$ & $  0.13^{+  0.01}_{-  0.01}$ & $  262.10$ \\
PTF09cwl    & $55.34^{+0.12}_{-0.14}$ & $-2.35^{+1.06}_{-1.04}$ & $ -0.24^{+  0.81}_{-  0.85}$ & $ 35.48^{+ 23.40}_{- 16.43}$ & $  9.67^{+  1.83}_{-  1.84}$ & $   33.15^{+   44.69}_{-   21.47}$ & $0.13^{+0.05}_{-0.04}$ & $  4.71^{+  1.55}_{-  1.03}$ & $  0.20^{+  0.18}_{-  0.13}$ & $  0.65^{+  0.11}_{-  0.11}$ & $   -9.39$ \\
PTF10hgi    & $54.49^{+0.04}_{-0.03}$ & $-0.32^{+0.34}_{-0.35}$ & $ -2.45^{+  0.27}_{-  0.26}$ & $  5.50^{+  3.02}_{-  1.51}$ & $  5.37^{+  0.14}_{-  0.20}$ & $    1.59^{+    1.00}_{-    0.52}$ & $0.12^{+0.04}_{-0.05}$ & $  6.43^{+  0.27}_{-  0.20}$ & $  0.06^{+  0.10}_{-  0.05}$ & $  0.16^{+  0.02}_{-  0.01}$ & $  167.33$ \\
PTF11rks    & $54.74^{+0.07}_{-0.09}$ & $0.90 ^{+0.25}_{-0.76}$ & $ -3.01^{+  0.58}_{-  0.26}$ & $  3.89^{+  1.12}_{-  0.80}$ & $ 11.18^{+  0.57}_{-  0.58}$ & $    4.87^{+    2.06}_{-    1.39}$ & $0.19^{+0.01}_{-0.02}$ & $  7.88^{+  0.33}_{-  0.42}$ & $  0.47^{+  0.02}_{-  0.07}$ & $  0.28^{+  0.02}_{-  0.02}$ & $  124.50$ \\
PTF12dam    & $55.28^{+0.02}_{-0.02}$ & $-0.61^{+0.60}_{-1.18}$ & $ -1.46^{+  0.81}_{-  0.42}$ & $ 33.88^{+ 22.35}_{-  8.18}$ & $  6.58^{+  0.26}_{-  0.29}$ & $   14.69^{+   11.68}_{-    4.50}$ & $0.13^{+0.04}_{-0.05}$ & $  9.98^{+  0.02}_{-  0.05}$ & $  0.01^{+  0.01}_{-  0.01}$ & $  0.60^{+  0.01}_{-  0.01}$ & $  -30.70$ \\
iPTF13ajg   & $55.44^{+0.05}_{-0.09}$ & $0.34 ^{+0.30}_{-1.09}$ & $ -1.94^{+  0.78}_{-  0.29}$ & $  7.76^{+  4.83}_{-  2.01}$ & $ 13.97^{+  1.22}_{-  0.52}$ & $   15.14^{+   13.88}_{-    4.74}$ & $0.17^{+0.02}_{-0.04}$ & $  4.73^{+  0.30}_{-  0.58}$ & $  0.47^{+  0.03}_{-  0.04}$ & $  0.14^{+  0.02}_{-  0.02}$ & $  132.27$ \\
iPTF13dcc   & $55.46^{+0.07}_{-0.05}$ & $-2.30^{+1.09}_{-1.07}$ & $ -0.16^{+  0.78}_{-  0.78}$ & $ 22.91^{+ 10.20}_{-  6.31}$ & $  5.29^{+  0.49}_{-  0.48}$ & $    6.42^{+    4.65}_{-    2.57}$ & $0.09^{+0.05}_{-0.03}$ & $  4.67^{+  0.34}_{-  0.29}$ & $  0.06^{+  0.10}_{-  0.04}$ & $  0.23^{+  0.02}_{-  0.03}$ & $   69.82$ \\
iPTF13ehe   & $55.39^{+0.03}_{-0.02}$ & $-2.21^{+1.14}_{-1.12}$ & $ -0.29^{+  0.78}_{-  0.78}$ & $ 77.62^{+ 11.50}_{- 10.02}$ & $  8.39^{+  0.23}_{-  0.37}$ & $   54.61^{+   11.64}_{-   11.19}$ & $0.16^{+0.03}_{-0.02}$ & $  5.00^{+  0.13}_{-  0.09}$ & $  0.04^{+  0.06}_{-  0.04}$ & $  0.06^{+  0.01}_{-  0.01}$ & $  116.75$ \\
iPTF16bad   & $54.45^{+0.03}_{-0.03}$ & $-1.17^{+0.89}_{-1.77}$ & $ -1.92^{+  1.21}_{-  0.62}$ & $  2.69^{+  1.11}_{-  0.55}$ & $  7.09^{+  0.47}_{-  0.29}$ & $    1.35^{+    0.82}_{-    0.36}$ & $0.08^{+0.03}_{-0.02}$ & $  5.69^{+  0.54}_{-  0.57}$ & $  0.05^{+  0.07}_{-  0.04}$ & $  0.07^{+  0.02}_{-  0.01}$ & $   99.17$ \\
PS1-10ahf   & $55.83^{+0.09}_{-0.12}$ & $1.96 ^{+0.03}_{-0.03}$ & $ -2.63^{+  0.11}_{-  0.14}$ & $  7.24^{+  3.72}_{-  2.00}$ & $  6.93^{+  0.80}_{-  1.20}$ & $    3.48^{+    3.06}_{-    1.76}$ & $0.17^{+0.02}_{-0.04}$ & $  7.78^{+  0.54}_{-  0.59}$ & $  0.30^{+  0.14}_{-  0.19}$ & $  0.19^{+  0.02}_{-  0.02}$ & $   81.79$ \\
PS1-10awh   & $54.91^{+0.07}_{-0.04}$ & $-1.06^{+0.87}_{-1.42}$ & $ -1.53^{+  1.02}_{-  0.62}$ & $  7.24^{+  4.24}_{-  2.00}$ & $ 10.79^{+  0.71}_{-  0.89}$ & $    8.43^{+    6.76}_{-    3.29}$ & $0.14^{+0.04}_{-0.06}$ & $  6.14^{+  1.02}_{-  1.29}$ & $  0.04^{+  0.06}_{-  0.03}$ & $  0.05^{+  0.01}_{-  0.01}$ & $  228.56$ \\
PS1-10bzj   & $54.87^{+0.06}_{-0.05}$ & $1.26 ^{+0.11}_{-0.15}$ & $ -3.12^{+  0.16}_{-  0.12}$ & $  1.70^{+  0.59}_{-  0.38}$ & $ 10.72^{+  0.55}_{-  0.56}$ & $    1.95^{+    0.96}_{-    0.59}$ & $0.15^{+0.03}_{-0.03}$ & $  6.74^{+  0.33}_{-  0.31}$ & $  0.18^{+  0.16}_{-  0.13}$ & $  0.14^{+  0.02}_{-  0.02}$ & $   87.90$ \\
PS1-10ky    & $55.13^{+0.05}_{-0.05}$ & $-0.86^{+1.08}_{-1.66}$ & $ -1.45^{+  1.16}_{-  0.77}$ & $  5.50^{+  2.82}_{-  1.42}$ & $ 10.30^{+  1.30}_{-  0.80}$ & $    5.83^{+    5.36}_{-    2.15}$ & $0.12^{+0.05}_{-0.04}$ & $  7.64^{+  0.45}_{-  0.30}$ & $  0.11^{+  0.15}_{-  0.07}$ & $  0.13^{+  0.02}_{-  0.02}$ & $  143.36$ \\
PS1-10pm    & $55.06^{+0.06}_{-0.04}$ & $-2.64^{+1.01}_{-1.03}$ & $ -0.33^{+  0.75}_{-  0.71}$ & $ 28.84^{+ 22.45}_{-  9.79}$ & $ 22.36^{+  1.51}_{-  2.11}$ & $  144.18^{+  148.07}_{-   66.06}$ & $0.15^{+0.04}_{-0.06}$ & $  7.40^{+  0.63}_{-  0.72}$ & $  0.23^{+  0.10}_{-  0.14}$ & $  0.09^{+  0.03}_{-  0.02}$ & $   53.64$ \\
PS1-11ap    & $55.53^{+0.06}_{-0.07}$ & $1.55 ^{+0.05}_{-0.09}$ & $ -2.65^{+  0.12}_{-  0.10}$ & $  5.62^{+  3.09}_{-  1.82}$ & $  5.98^{+  0.27}_{-  0.34}$ & $    2.01^{+    1.39}_{-    0.80}$ & $0.11^{+0.05}_{-0.04}$ & $  7.62^{+  0.53}_{-  1.55}$ & $  0.34^{+  0.13}_{-  0.16}$ & $  0.28^{+  0.01}_{-  0.01}$ & $  265.23$ \\
PS1-11bam   & $55.25^{+0.13}_{-0.16}$ & $-0.38^{+1.61}_{-2.34}$ & $ -1.65^{+  1.69}_{-  1.23}$ & $  5.01^{+  5.22}_{-  1.78}$ & $  8.67^{+  0.94}_{-  0.81}$ & $    3.77^{+    5.69}_{-    1.77}$ & $0.10^{+0.07}_{-0.04}$ & $  5.91^{+  0.94}_{-  1.08}$ & $  0.08^{+  0.10}_{-  0.06}$ & $  0.09^{+  0.03}_{-  0.03}$ & $   69.73$ \\
PS1-14bj    & $55.53^{+0.06}_{-0.12}$ & $1.80 ^{+0.17}_{-0.65}$ & $ -2.82^{+  0.49}_{-  0.23}$ & $ 32.36^{+ 11.29}_{-  7.24}$ & $  3.39^{+  0.37}_{-  0.34}$ & $    3.72^{+    2.44}_{-    1.38}$ & $0.17^{+0.02}_{-0.03}$ & $  9.14^{+  0.35}_{-  0.29}$ & $  0.06^{+  0.08}_{-  0.04}$ & $  0.23^{+  0.03}_{-  0.03}$ & $   56.58$ \\
LSQ12dlf    & $54.99^{+0.04}_{-0.03}$ & $0.70 ^{+0.27}_{-0.42}$ & $ -2.63^{+  0.32}_{-  0.21}$ & $  3.80^{+  1.82}_{-  0.92}$ & $  8.47^{+  0.22}_{-  0.31}$ & $    2.73^{+    1.52}_{-    0.81}$ & $0.13^{+0.04}_{-0.04}$ & $  3.59^{+  0.13}_{-  0.12}$ & $  0.24^{+  0.11}_{-  0.14}$ & $  0.08^{+  0.01}_{-  0.01}$ & $  145.27$ \\
LSQ14mo     & $54.40^{+0.04}_{-0.05}$ & $-2.52^{+0.93}_{-1.14}$ & $ -1.07^{+  0.80}_{-  0.67}$ & $  6.92^{+  4.56}_{-  1.67}$ & $ 14.06^{+  0.46}_{-  0.48}$ & $   13.68^{+   10.54}_{-    4.00}$ & $0.14^{+0.04}_{-0.06}$ & $  4.71^{+  0.09}_{-  0.11}$ & $  0.16^{+  0.08}_{-  0.08}$ & $  0.02^{+  0.02}_{-  0.02}$ & $  158.16$ \\
LSQ14bdq    & $55.33^{+0.11}_{-0.05}$ & $-2.70^{+1.12}_{-0.99}$ & $ -0.02^{+  0.77}_{-  0.80}$ & $ 52.48^{+ 21.65}_{- 14.46}$ & $ 12.12^{+  2.20}_{-  1.66}$ & $   77.13^{+   74.98}_{-   35.48}$ & $0.15^{+0.04}_{-0.05}$ & $  6.12^{+  0.81}_{-  0.53}$ & $  0.15^{+  0.20}_{-  0.11}$ & $  0.37^{+  0.05}_{-  0.05}$ & $   14.49$ \\
Gaia16apd   & $55.01^{+0.04}_{-0.03}$ & $0.22 ^{+0.25}_{-0.51}$ & $ -2.29^{+  0.38}_{-  0.20}$ & $  5.75^{+  3.16}_{-  1.39}$ & $  9.13^{+  0.27}_{-  0.35}$ & $    4.80^{+    3.09}_{-    1.44}$ & $0.15^{+0.04}_{-0.06}$ & $  8.33^{+  0.21}_{-  0.19}$ & $  0.09^{+  0.05}_{-  0.05}$ & $  0.21^{+  0.01}_{-  0.01}$ & $  379.84$ \\
DES14X3taz  & $55.01^{+0.04}_{-0.08}$ & $-2.11^{+0.87}_{-1.12}$ & $ -0.73^{+  0.79}_{-  0.66}$ & $ 16.60^{+ 12.92}_{-  5.63}$ & $ 12.66^{+  1.79}_{-  1.65}$ & $   26.60^{+   35.06}_{-   13.30}$ & $0.13^{+0.05}_{-0.06}$ & $  5.95^{+  0.36}_{-  0.34}$ & $  0.41^{+  0.07}_{-  0.13}$ & $  0.28^{+  0.04}_{-  0.03}$ & $   29.87$ \\
SCP-06F6    & $55.73^{+0.16}_{-0.11}$ & $1.69 ^{+0.11}_{-0.12}$ & $ -2.55^{+  0.24}_{-  0.18}$ & $  7.76^{+  3.20}_{-  2.51}$ & $  7.15^{+  1.27}_{-  1.04}$ & $    3.97^{+    3.80}_{-    2.01}$ & $0.15^{+0.04}_{-0.04}$ & $  5.84^{+  1.10}_{-  1.03}$ & $  0.24^{+  0.15}_{-  0.17}$ & $  0.60^{+  0.17}_{-  0.11}$ & $  -10.58$ \\
SNLS06D4eu  & $55.18^{+0.07}_{-0.07}$ & $1.29 ^{+0.11}_{-0.15}$ & $ -2.83^{+  0.17}_{-  0.14}$ & $  2.40^{+  0.69}_{-  0.74}$ & $ 11.93^{+  0.97}_{-  0.56}$ & $    3.42^{+    1.73}_{-    1.27}$ & $0.18^{+0.02}_{-0.04}$ & $  6.12^{+  0.91}_{-  1.13}$ & $  0.09^{+  0.10}_{-  0.06}$ & $  0.14^{+  0.02}_{-  0.02}$ & $   52.85$ \\
SNLS07D2bv  & $54.90^{+0.05}_{-0.04}$ & $-1.28^{+0.27}_{-1.40}$ & $ -1.40^{+  0.98}_{-  0.22}$ & $ 21.38^{+ 10.98}_{-  5.16}$ & $ 14.52^{+  0.56}_{-  0.60}$ & $   45.11^{+   28.58}_{-   13.67}$ & $0.15^{+0.03}_{-0.04}$ & $  6.12^{+  0.85}_{-  1.16}$ & $  0.25^{+  0.06}_{-  0.08}$ & $  0.07^{+  0.01}_{-  0.00}$ & $  156.92$ \\
SSS120810   & $54.64^{+0.04}_{-0.05}$ & $-1.34^{+1.56}_{-1.59}$ & $ -1.62^{+  1.10}_{-  1.09}$ & $  2.88^{+  1.69}_{-  0.75}$ & $ 11.01^{+  1.12}_{-  0.81}$ & $    3.49^{+    3.23}_{-    1.27}$ & $0.12^{+0.06}_{-0.04}$ & $  3.49^{+  0.11}_{-  0.11}$ & $  0.23^{+  0.20}_{-  0.12}$ & $  0.21^{+  0.05}_{-  0.03}$ & $   28.85$ \\
\enddata
\end{deluxetable*}
\end{longrotatetable}






\bibliography{references}



\end{document}